\documentclass[twocolumn]{aa}

\usepackage{graphicx}
\usepackage{url}
\usepackage[colorlinks=true]{hyperref}
\hypersetup{citecolor= blue, linkcolor=red, urlcolor=blue}
\usepackage{txfonts}
\usepackage{float}%,biblatex}
\usepackage{natbib}
\bibpunct{(}{)}{;}{a}{}{,} % to follow the A&A style
\usepackage[us,12hr]{datetime}
\usepackage{ulem}
\usepackage{cancel}

\begin{document} 

   \title{Solar image denoising with convolutional neural networks}

   % \subtitle{I. Aspectos interesantes}

   \author{C. J. D\'iaz Baso
          \inst{1}
          \and
          J. de la Cruz Rodr\'iguez
          \inst{1}
          \and
          S. Danilovic
          \inst{1}
          }

   \institute{Institute for Solar Physics, Dept. of Astronomy, Stockholm University, AlbaNova University Centre, SE-10691 Stockholm Sweden
              \email{carlos.diaz@astro.su.se}
             }

    \date{Received June 11, 2019; accepted August 7, 2019}
   % \date{Draft: compiled on \today\ at \currenttime~UT}
   
   \authorrunning{D\'iaz Baso et al.}
   % \titlerunning{Magnetic field of an active region filament}
% \abstract{}{}{}{}{} 
% 5 {} token are mandatory
 
  \abstract{The topology and dynamics of the solar chromosphere are greatly affected by the presence of magnetic fields. The magnetic field can be inferred by analyzing polarimetric observations of spectral lines. Polarimetric signals induced by chromospheric magnetic fields are, however, particularly weak, and in most cases very close to the detection limit of current instrumentation. Because of this, there are only few observational studies that have successfully reconstructed the three components of the magnetic field vector in the chromosphere. Traditionally, the signal-to-noise ratio of observations has been improved by performing time-averages or spatial averages, but in both cases, some information is lost. More advanced techniques, like principal-component-analysis, have also been employed to take advantage of the sparsity of the observations in the spectral direction. In the present study, we propose to use the spatial coherence of the observations to reduce the noise using deep-learning techniques. We design a neural network that is capable of recovering weak signals under a complex noise corruption (including instrumental artifacts and non-linear post-processing). The training of the network is carried out without a priori knowledge of the clean signals, or an explicit statistical characterization of the noise or other corruption. We only use the same observations as our generative model. The performance of this method is demonstrated on both, synthetic experiments and real data. We show examples of the improvement in typical signals obtained in current telescopes such as the Swedish 1-meter Solar Telescope. The presented method can  recover weak signals equally well no matter on what spectral line or spectral sampling is used. It is especially suitable for cases when the wavelength sampling is scarce.}

%   Finally we show how the inference of the signal improves if the network takes into account not only the spatial information but also the spectral information.}
  % % aims heading (mandatory)
  %  {...}
  % % methods heading (mandatory)
  %  {...}
  % % results heading (mandatory)
  %  % {...}
  %  {...}
  % % conclusions heading (optional), leave it empty if necessary 
  %  {}


   \maketitle

% https://www.aanda.org/index2.php?option=com_content&task=view&id=170&Itemid=256

%%%%%%%%%%%%%%%%%%%%%%%%%%%%%%%%%%%%%%%%%%%%%%%%%%%%%%%%%%%%%%%%%%%%%%%%%
\section{Introduction}\label{sec:intro}
%%%%%%%%%%%%%%%%%%%%%%%%%%%%%%%%%%%%%%%%%%%%%%%%%%%%%%%%%%%%%%%%%%%%%%%%%

The magnetic field plays a key role in the generation and evolution of many of the phenomena that take place in the Sun, from events with sizes that are invisible to our largest telescopes to spectacular large high energy eruptions \citep{Wiegelmann2014}, and its study is thus mandatory. The information about the magnetic fields is encoded in the polarization of the radiation of the Sun and the stars through the Zeeman and Hanle effects \cite[e.g.,][]{Landi_landolfi04,TrujilloBueno2010}. Stokes polarimetry is, therefore, the observation tool that gives us access to this information and allows us to determine the topology and evolution of the magnetic field. At photospheric levels outside active regions, as well as in most of the chromosphere and corona, magnetic fields are weaker and, therefore the polarization signals are often close to the noise level \citep[e.g.,][]{JaimeReview2017}. This has led to very challenging requirements in the instrumentation of current telescopes as well as the new era of large telescopes, such as the 4-m Daniel K. Inouye Solar Telescope \cite[DKIST;][]{DKIST2016} or the 4-m European Solar Telescope \cite[EST;][]{EST2013}.
 
For a correct analysis of polarization signals, it is necessary to identify and remove instrumental artifacts from the detected signals. One of the most recurrent effects in spectropolarimetric data is the appearance of interference patterns commonly known as fringes \citep{Lites1991,Semel2003,Harrington2017}. Several studies have successfully identified and removed them from the observed signals using Fourier filtering, wavelet analysis \citep{Rojo2006} and different implementations of Principal Component Analysis \cite[PCA;][]{Pca1963,Casini2018}. However, there are other issues that are not as simple to mitigate, such as the intrinsic noise associated with measurements. The characteristics of this noise depend on multiple factors: the source we are studying, the sensitivity of the detectors, and the modification of the signal during the exposure, storage, transmission, processing, and conversion. The most natural way to increase the signal to noise ratio (S/N) is to increase the exposure time or perform a time average. However, this procedure can be critical for events that change very rapidly in time (e.g. in the solar chromosphere) and therefore important information is lost by performing this operation. Another approach can be simply to perform a spatial average of some pixels, something that also removes spatial information of the distribution of the signals. The most widely used procedure in recent years to improve signals without losing information has been the application of PCA as a denoising technique \citep{Carroll2007,Marian2008}. With this technique, we assume that there is a small space of vectors that can explain most of the spectral profiles of our field of view (FOV) and that all of them can be decomposed as a linear combination of the elements of this basis. With PCA we are able to find the new representation (a basis of vectors) that best reproduces the observations and that the large variance is explained by the first vectors of the basis. Assuming that the noise is uncorrelated, denoising is possible by reconstructing the data with a truncated basis.  This technique is based on the spectral information present and the correlation between the measured wavelength points. Therefore, if wavelength sampling is scarce, the efficiency of PCA reconstruction decreases.

In this study we explore, through the use of a neural network, the idea of using the presence of spatial correlation, properties such as smoothness and structures in the image, to predict the value of a pixel from the value of other pixels. The idea can be exploited by  with convolutional neural networks \citep[CNN;][]{LeCun1998}. They are composed of several convolutional neurons, where each CNN-neuron carries out the convolution of the input with a certain {kernel}. The output is known as feature map and contains the information on how each pixel relates to its neighbors. CNNs have been successfully used in many application in the solar physics field: to infer horizontal velocity fields from consecutive continuum images \citep{DeepVel2017}, for solar flare prediction \citep{DeepFlare2018,DeepFlare2018b}, to efficiently deconvolve solar images \citep{Enhance2018,DeepMOMFBD2018}, for coronal holes segmentation \citep{Illarionov2018}, for spectropolarimetric inversions \citep{2019arXiv190403714A,Osborne2019} and many others that are being developed \citep{monica_bobra_2019_2575738}.

Among different convolutional neural network approaches for denoising, probably the best example are the \emph{autoencoder networks}. This type tries to reconstruct the image from its corrupted version by transforming it into another representation, usually a low-dimensional space, also called latent space. There, statistical regularities are more easily captured, after which the corrected image is re-projected back in the input space \citep{Vincent2008,Jain2008}. Based on the same idea of reducing data dimensionality as PCA, neural networks are much more flexible as this transformation is highly non-linear. Similar approaches (and also other architectures) have made it possible to develop neural networks that can denoise an image with an unknown noise level, or even a spatially variant noise \citep{Zhang2017,Zhang2018,Ehret2018}. To recover the underlying image, these neural networks are trained with artificial noise and therefore prior information about the noise, like the type, expected level or the spatial distribution, is required. In fact, the assumption of white Gaussian or Poisson noise (justified by the photon count process at the image sensor) is not valid anymore because we usually do not have access to raw data, but processed data where the noise is spatially correlated and signal dependent due to the application of image reconstruction techniques (e.g., MOMFBD, \citealt{MFBD2002}; \citealt{vanNoort2005}). Moreover, these characteristics depend on both the instrument and the processing pipeline. Therefore, the characterization of noise is not always trivial in solar applications.

In contrast, the information from a single image and the ability of neural networks to generate natural images have shown to be sufficient for a high-quality correction \citep{DeepPrior2017}. In these studies, a randomly initialized convolutional network resists generating noisy images and therefore succeeds in correcting the image (among other tasks such as superresolution) from a single image. The problem with this technique is that although clean data is not needed a priori, it is a very slow and computationally expensive technique because it needs to generate a network for each image.
%http://mlexplained.com/2018/01/18/paper-dissected-deep-image-prior-explained/ 

Although all the previous techniques have been very useful to show the capabilities of the neural networks as a denoiser, they present some disadvantages. Our analysis is inspired by Noise2Noise \citep{Noise2Noise2018} approach which is more suitable for the problem for two reasons.  First, the method uses certain properties of neural networks to clean corrupted data, without need to have the pair noisy-clean image, i.e. without the need to generate data with synthetic noise.  Second, there is also no need for an explicit statistical probability model of the noise corruption or a clean image, because the network learns it indirectly from the data. In fact, the authors show how to transform corrupted into clean images just by looking at bad images, and do it as well (sometimes even better) as if they used clean examples.

Following this idea, we propose how to apply this technique to spectro-polarimetric images in order to reconstruct weak signals present in the observations. As the neural network learns from corrupted images, this technique can be applied to any telescope, instrumentation, processing pipeline, solar region or physical observable.

The paper is organized as follows: first we explain how we implement the Noise2Noise approach with spectro-polarimetric data, then we demonstrate that the technique works on synthetic data and finally we apply it to the real solar data.

%%%%%%%%%%%%%%%%%%%%%%%%%%%%%%%%%%%%%%%%%%%%%%%%%%%%%%%%%%%%%%%%%%%%%%%%%
\section{Neural network denoising}\label{sec:Denoising}
%%%%%%%%%%%%%%%%%%%%%%%%%%%%%%%%%%%%%%%%%%%%%%%%%%%%%%%%%%%%%%%%%%%%%%%%%

%%%%%%%%%%%%%%%%%%%%%%%%%%%%%%%%%%%%%%%%%%%%%%%%%%%%%%%%%%%%%%%%%%%%%%%%%
\subsection{Training process}\label{sec:training}
%%%%%%%%%%%%%%%%%%%%%%%%%%%%%%%%%%%%%%%%%%%%%%%%%%%%%%%%%%%%%%%%%%%%%%%%%

\begin{figure}[!ht]
\centering
\includegraphics[width=\linewidth]{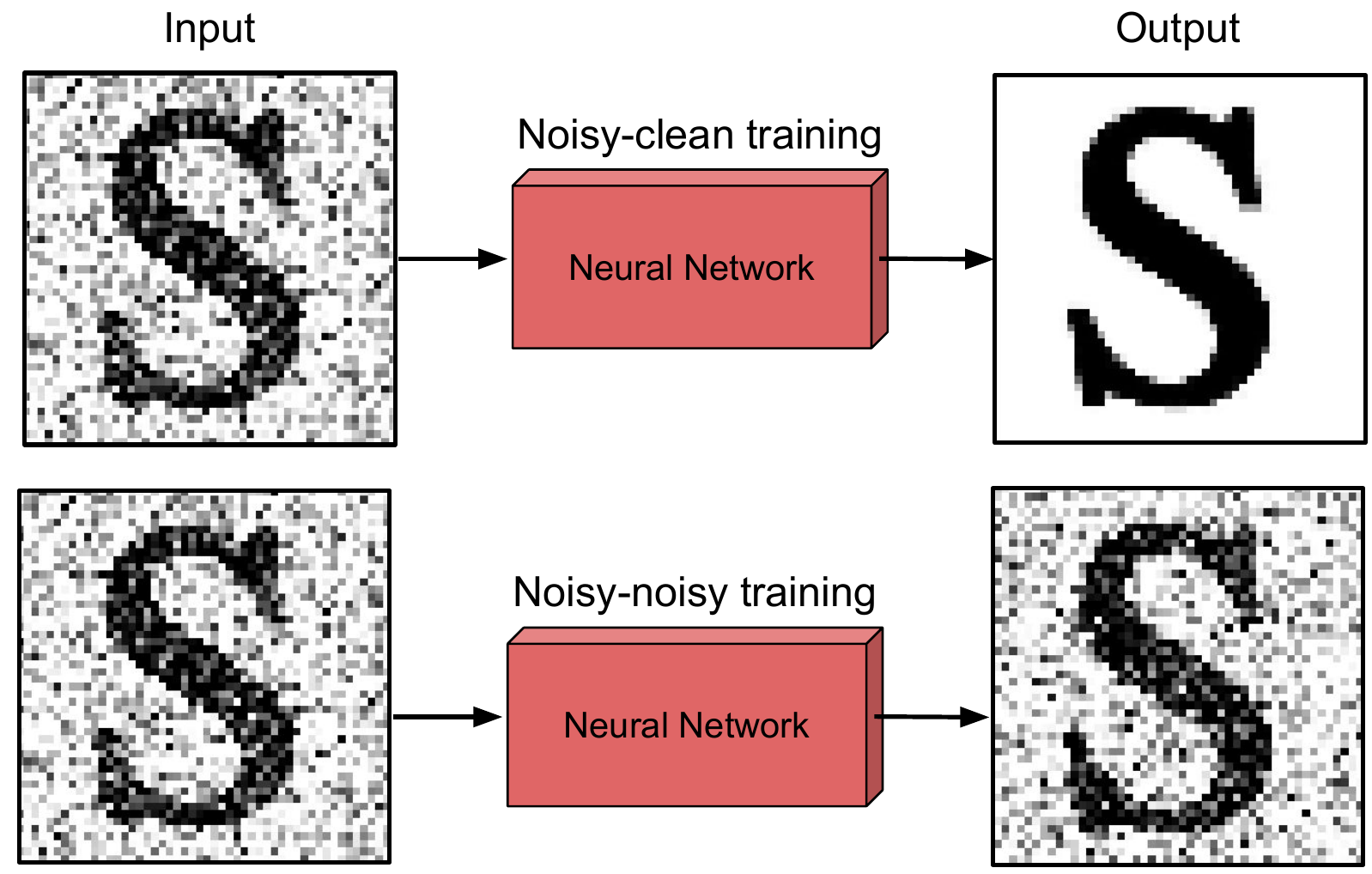}
\caption{Sketch of different training processes: the typical supervised training process where the clean target (answer of the problem) is used (upper panel) and, when different uncorrelated corrupted images are used as the input and output of the network (lower panel).}
\label{fig:training}
% GOOGLE_DRIVE
\end{figure}

Prior to the study by \cite{Noise2Noise2018}, it was common to train a regression model with $\theta$ parameters, such as a convolutional neural network, with a large number of pairs $(x_i,y_i)$ of corrupt inputs and clean targets. The parametric mapping $f_\theta$ was then optimized under a merit function that minimizes the distance between the target and the neural network output. The obtained $f_\theta(x)$ was finally applied and the noise of corrupted images reduced. This process, where we would know the correct answer (supervised learning), is schematically represented in the upper panel of Fig.~\ref{fig:training}. 

\cite{Noise2Noise2018} demonstrated that in case of data corrupted with zero-mean noise, the neural network can be trained with a large number of pairs $(x_i,y’_i)$ of corrupted inputs and targets, corresponding to the same underlying data and independent noise realizations, thus eliminating the need for clean data. This process is illustrated in the lower panel of Fig.~\ref{fig:training}. This is of great relevance since in many cases obtaining clean training targets or creating synthetic examples with correct noise modeling is very difficult or impossible. That is exactly the case of solar spectropolarimetric measurements, where not only the noise is a fundamental factor, but the images are also affected by the telescope and the image reconstruction. Unlike the first case, the neural network derives insights in an unsupervised way directly from the data itself. From the pairs of independent noisy observations of the same underlying clean image, the network can retrieve the information about the noise. We take advantage of, for example, the temporal redundancy in time series measurements, which means, the noise is the main difference between the two images. %https://arxiv.org/pdf/1812.07715v1.pdf

In the following, we show the implementation of this new training technique to spectropolarimetric data, as well as the neural network architecture and the optimization process. For many of the technical details, we refer the reader to our previous work \citep{Enhance2018} that contains an elaborate introduction about neural networks. We explain in that work the motivation of using each type of layer and the fundamentals of the training process.

%%%%%%%%%%%%%%%%%%%%%%%%%%%%%%%%%%%%%%%%%%%%%%%%%%%%%%%%%%%%%%%%%%%%%%%%%
\subsection{Topology and optimization}\label{sec:topology}
%%%%%%%%%%%%%%%%%%%%%%%%%%%%%%%%%%%%%%%%%%%%%%%%%%%%%%%%%%%%%%%%%%%%%%%%%

\begin{figure}[!ht]
\centering
\includegraphics[width=\linewidth]{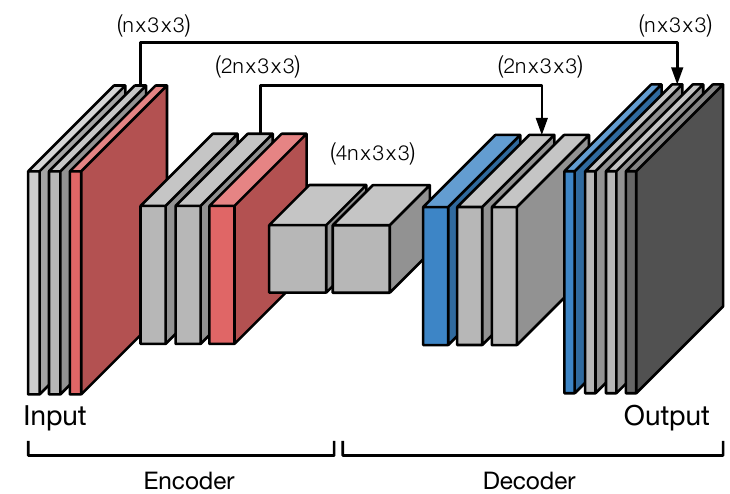}
\caption{Architecture of the convolutional network used for denoising, made of many Conv+ELU gray  blocks and the concatenation (arrows) of previous feature maps. In an autoencoder, the information is compressed {(encoder) and recovered later (decoder)}.}
\label{fig:topology}
% Drive
\end{figure}

The technique is implemented using an encoder-decoder architecture. A schematic view of it is displayed in Fig.~\ref{fig:topology}. This network is very fast and has demonstrated to be able to beat the best previous methods by using only very few images \citep{Unet2015}. It decreases and increases the dimensions of feature maps to capture features at different image scales. In the encoder phase, the spatial size of images is reduced and in the decoder phase, the original size is recovered by upsampling.

This specific topology is commonly known as U-network\footnote{A similar implementation in \texttt{keras} can be found in \url{https://github.com/pietz/unet-keras}} \citep{Unet2015} because it includes skip connections between mirrored layers in the encoder and decoder part. Following the improvements of \cite{Huang2016Densely}, it does not combine features through summation before they are passed into a layer; instead, it combines features by concatenating them (arrows in Fig.~\ref{fig:topology}). In the encoder part of the network \footnote{{A detailed explanation of the architecture of the network and the number of filters in each block is given in Appendix \ref{tab:arquitecture}}}, the information is transferred between gray blocks and a max-pooling layer (red block) that reduces the size by half. Each gray block consists of a convolution layer with $n$ kernel filters (of size 3$\times$3) and ReLU activating function \citep{relu10}. To preserve the amount of information, the number of feature maps (filters) is doubled after each max-pooling layer (quoted above each block of Fig.~\ref{fig:topology}). All convolutions use a reflection padding to keep the original size and avoid border effects. In the decoder part, we have implemented a nearest-neighbor upsampling layer followed by a convolution layer (highlighted as blue blocks in Fig.~\ref{fig:topology}) instead of the standard transpose convolution to avoid checkerboard artifacts in the upsampling\footnote{\url{https://distill.pub/2016/deconv-checkerboard/}.}. The last layer, drawn as a dark-gray block, is a 1$\times$1 convolution layer that quickly collapses the previous information into the output.

After extensive tests, we have verified that this topology is the perfect balance between accuracy and speed of execution. {We take as the best neural network model the one whose inference using the validation set is closer to the expected output. The speed of execution} is of great importance since the network will be applied to each wavelength independently for each Stokes parameter and for each frame of the time series. Then, the dimension of the input and the output is always a two-dimensional image. Other architectures are also studied in section \ref{sec:constrain}.
 
Although this complex neural network\footnote{The code of the network and the weights obtained after the training can be found in \url{https://github.com/cdiazbas/denoiser/}.} is able to recreate the results we show, simpler implementations can lead to similar results but not as accurate. An example of this is the neural network presented in \cite{Enhance2018} (without the last upsampling layer). It is a network that keeps the size of the image throughout its topology. This architecture has been used before also as a denoiser in recent studies \citep{Zhang2017,Zhang2018,Ehret2018,Mansar2018}. While its composition is similar, the U-network topology has been designed to reduce dimensionality and efficiently capture features at different scales.

\begin{figure*}[!ht]
\centering
\includegraphics[width=\linewidth]{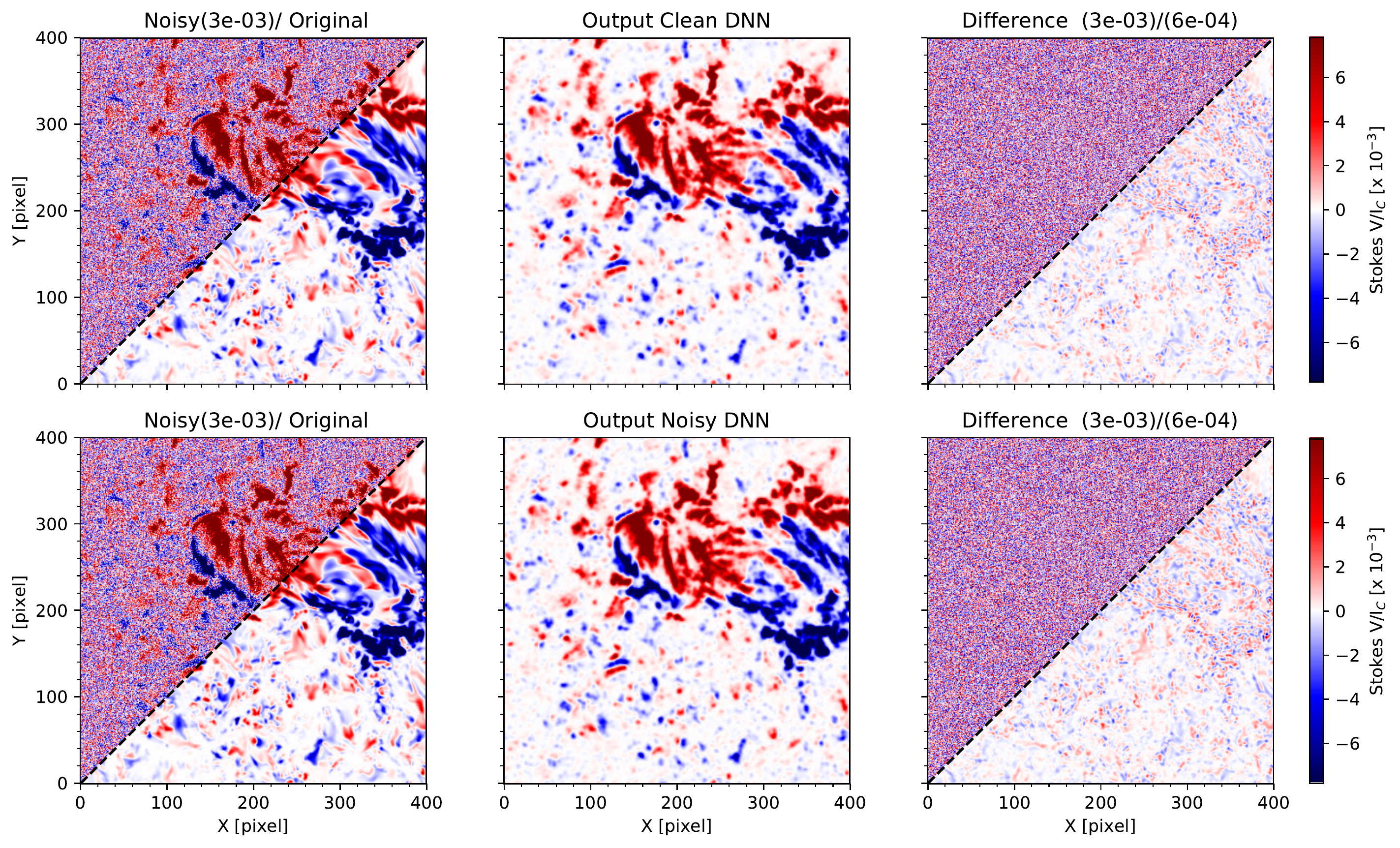}
\caption{Result of the reconstructions of synthetic data with an original noise level of 3$\times10^{-3}I_c$, using clean targets (upper panel) and using noisy targets (lower panel). The left column shows the original image (and affected by noise in the upper half), the central panel the image inferred by the neural network and the right panel the difference between the previous panels of the same row.}
\label{fig:synthetic}
% /scratch/carlos/DEEPL/DENOISEPAPER/final/simulacion/plotjuntos.py
\end{figure*}

The neural network is trained to optimize a merit function that measures how different is the network output when compared to target images. In the following, we first show the results using this topology and training the network with the merit function $L_2=1/N\sum_i^N (f_\theta(x_i)-y_i)^2$, i.e. the quadratic difference between the output data and the target.  If the data is affected with a significant (up to 50\%) outlier content, the $L_1=1/N\sum_i^N |f_\theta(x_i)-y_i|$ the merit function would be more suitable, making outliers less important. We have tested both and the differences were very small, with $L_2$ giving better results. A different merit function is used for estimating the uncertainty of the inference in section \ref{sec:uncertainty}.

The neural networks are optimized with the Adam stochastic gradient descent type algorithm \citep{adam14} with a learning rate of $10^{-4}$ that was kept constant during the whole training. In all the examples we have kept a very low or non-existent weight regularization. Although an increase in it generates smoother structures (more spatial coherence), we lose the ability to resolve details. This extra regularization is not really necessary since both input and output already contain the necessary noise to regularize the weights and the behavior of the network. No batch normalization layers were used. The network weights were initialized following \cite{He2015}. 

After finding a topology that responds successfully to our objective, we studied whether increasing the number of filters (and therefore feature maps) could improve the accuracy of the network. For this, we have trained models with different amounts of filters at the entrance of the network and we have found that $n=$32 filters give us a good balance between accuracy and speed. 
%64 = 1e6, 32 = 5e5, 16 = 1e5, 8 = 5e4

Finally, the typical training strategy depends on the data set and the instrumentation used. Therefore in each section, we give details of the specific procedure that was carried out.

%%%%%%%%%%%%%%%%%%%%%%%%%%%%%%%%%%%%%%%%%%%%%%%%%%%%%%%%%%%%%%%%%%%%%%%%%
\section{Results}\label{sec:results}
%%%%%%%%%%%%%%%%%%%%%%%%%%%%%%%%%%%%%%%%%%%%%%%%%%%%%%%%%%%%%%%%%%%%%%%%%

This section shows how the technique performs in both a supervised environment (synthetic data) and in real observations. In these examples, the only difference is the data training, while the network architecture and the rest of the parameters remain the same. The training is done with images of size 52$\times$52 in both cases. This size is chosen arbitrarily so it is large enough to capture the scale of the noise and not be affected by border effects during convolutions. Larger patches can also be chosen, but this would slow down the training process. %The calculation time is in the order of 1\,s per image (1k$\times$1k) using a CPU.

%%%%%%%%%%%%%%%%%%%%%%%%%%%%%%%%%%%%%%%%%%%%%%%%%%%%%%%%%%%%%%%%%%%%%%%%%
\subsection{Simulated data set}\label{sec:simulation}
%%%%%%%%%%%%%%%%%%%%%%%%%%%%%%%%%%%%%%%%%%%%%%%%%%%%%%%%%%%%%%%%%%%%%%%%%

In order to demonstrate the capabilities of this denoising technique, we have generated a synthetic case as proof of concept. With this test, we can assess which effects are due to the intrinsic problem of image restoration and which are due to this new training technique. We use an MHD simulation of a flux emergence produced with MURaM code \citep{2005A&A...429..335V,2017ApJ...834...10R}. The dimensions of the computational domain are $24\times 12 \times 8.2$~Mm with the grid spacing of 23 and 16~km in horizontal and vertical direction, respectively. The emergence of a strongly twisted bipolar magnetic concentration is imposed at the bottom boundary in the same way as in \cite{2019NatAs...3..160C}. The horizontal magnetic field is advected within an ellipsoidal region with axes of 6 and 1.5 Mm through the bottom boundary set at 1.5 Mm below photosphere. The chosen snapshot is taken some 1.5~h after, at which point the total amount of horizontal flux advected in is $6.9 \times 10^{19}$~Mx. The snapshot covers a variety of solar conditions from quiet Sun in the external parts to high magnetic field concentrations in the center of the FOV. Since our denoising technique does not depend on the spectral line or size of the magnetic features detected, we choose to synthesize  the \ion{Fe}{i} line at 6302~\AA . The polarized spectrum of the \ion{Fe}{i} line at 6302~\AA\  is synthesized over the wavelength range of 1~\AA\ with a spectral resolution of 50 m\AA. The synthetic spectrum emergent from this numerical simulation has been calculated under the assumption of local thermodynamic equilibrium (LTE). To simulate a validation/test set slightly different from the training, we have synthesized the \ion{Fe}{i} line at 6301~\AA\ in LTE using the STiC code \citep{2016ApJ...830L..30D,2019A&A...623A..74D}, to check the generalization of the network and to test the network. As the \ion{Fe}{i} line at 6302~\AA\ has a larger Land\'e factor, the training set contains a larger range of signals.

To compare the accuracy of this new technique with respect to the standard noisy-clean method, we have trained two neural networks, one with noisy-clean pairs and the other with noisy-noisy pairs. Both networks are trained during 20 epochs. {We have found that 20 epochs are sufficient for the network to reach the optimal state. As our network is not able to memorize random noise (independent in each batch), the merit function achieves a constant asymptotic tendency, without overfitting in case more epochs are used}. We have used a batch size of 10 images using a total of N = 10000 patches of 52$\times$52 pixels randomly extracted from the whole simulation. They are also randomly extracted from the spectral positions and from different Stokes parameter ($Q$, $U$ and $V$). The validation consists on 1000 patches of the same size from the other spectral line. The images have been trained adding Gaussian noise with a standard deviation of 3$\times10^{-3}I_c$. Figure \ref{fig:synthetic} shows the result of both reconstructions, using clean targets (upper row) and using noisy targets (lower row) {of a monochromatic image of Stokes $V$ at 6301.4\,\AA\ over a part of the simulated field of view}.

In the training process, both models converge similarly quickly and the denoising performance achieves almost the same precision. A proof of the latter is shown in the last column of Fig.~\ref{fig:synthetic}, where both neural networks succeed in correcting up to a level of $6\times10^{-4}I_c$. This value has been calculated as the standard deviation between the original clean image and the one inferred by the neural network. This means that we would need to average around 25 images ($3\times10^{-4}I_c\cdot\sqrt{25}=6\times10^{-4}I_c$) to reduce the noise achieved by the neural network.

Another confirmation that the neural network is recovering signal correctly is the fact that when we subtract the reconstructed from the noisy image we  obtain the same Gaussian distribution with a standard deviation of 3$\times10^{-3}I_c$, which we added to the synthetic image. Finally, there are no clear patterns that show that the network works much better in one place than in another. With this example we have shown that using clean targets is not necessary for this application and that real observations can be used to train a neural network.

\begin{figure}[!ht]
\centering
\includegraphics[width=\linewidth]{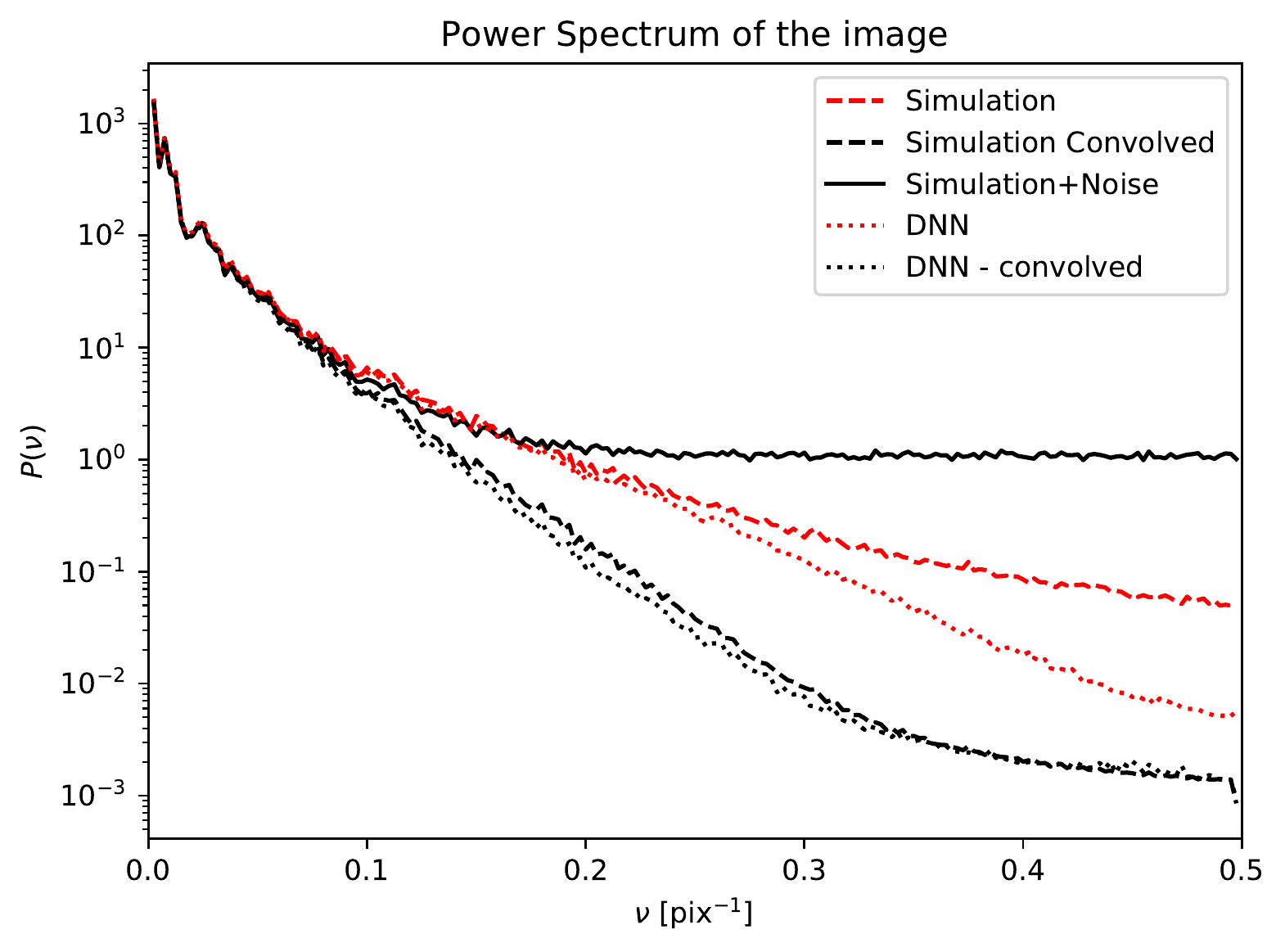}
\caption{Comparison of the power spectrum calculated for the simulation and the reconstruction with the neural network trained with the original or a degraded resolution.}
\label{fig:vpowersimu}
% /scratch/carlos/DEEPL/DENOISEPAPER/final/simulation/powerplot
\end{figure}

% \comment{Parrafo re-escrito entero}\\
When the denoising method is applied to an image, both the spatial scale and the amplitude of the signal affect the output of the NN. In situations where the spatial scale of a feature is close to the smallest scale allowed by our sampling and its amplitude is close to the noise level, the network will struggle to differ noise from real signals. When the data are oversampled, the larger spatial extent of the same feature in pixels allows the network to better differentiate noise from signal. We have illustrated these two situations in Fig.~\ref{fig:vpowersimu}. 

The red curve depicts the critically sampled case, whereas the black curve depicts an oversampled case. In the latter, we basically decreased the resolution of the data by applying a spatial Gaussian filter with a full width half maximum of 2 pixel before adding the noise to the synthetic data. This degradation effectively makes the image oversampled in the original pixel grid. In the oversampled case, most of the detail is recovered after the denoising step and both dashed and dotted lines closely overlap with each other. In the critically sampled case, there is a clear suppression of the highest spatial frequencies.

In this section, we have generated uncorrelated spatial noise, but in real observations, due to instrumentation and the post-processing, this quantity is highly correlated. That is, however, not a problem for the neural network as already shown by \cite{Noise2Noise2018} and \cite{Ehret2018}. It is possible to deduce the noise from the training data and remove its contribution whatever its complexity: uniform, multiplicative, correlated or salt-and-pepper. An example of complex corruption can be seen in the results with real observations.

%%%%%%%%%%%%%%%%%%%%%%%%%%%%%%%%%%%%%%%%%%%%%%%%%%%%%%%%%%%%%%%%%%%%%%%%%
\subsubsection{Uncertainty in the inference}\label{sec:uncertainty}
%%%%%%%%%%%%%%%%%%%%%%%%%%%%%%%%%%%%%%%%%%%%%%%%%%%%%%%%%%%%%%%%%%%%%%%%%

After demonstrating that it is possible to obtain a good estimation of the clean image, it is also necessary to have at least an idea of the uncertainty of our result, something that in principle this simple implementation does not capture. This information is really important for the later magnetic field inference and its respective uncertainty according to the reconstruction of the neural network.

Traditionally, Bayesian statistics have been used to infer the uncertainty for simple forward models with a reasonable number of parameters, usually less than 10 \citep[e.g.][]{Asensio2007}, as these calculations usually come with a prohibitive computational cost. However, our neural network model has a several orders of magnitudes more parameters than the previous models. For that reason, several studies \citep[among them,][]{Gal2015,Gal2017,Kendall2017} have shown a practical technique to estimate the total uncertainty. Uncertainty is usually classified into two categories: the epistemic and the aleatoric uncertainty. 

Epistemic uncertainty is related to unexplored regions of the mapping space and it is often called model uncertainty because it give us information how well the network learns about the data. Epistemic uncertainty can be estimated using the dropout technique (\citealt{Hinton2012}, hereafter dropout). In this technique a certain fraction of neurons are randomly deactivated when evaluating the network. Many modern models use this technique to avoid over-fitting during training. However, it can also be used for testing in which case the output is evaluated by dropping weights randomly and generating different predictions. 
%
%This is used to approximate the predictive distribution and measure the robustness of the neural network to a given prediction. 
%
So using this method, we can also assess the robustness of the predictions given by the neural network.
After performing these Monte Carlo calculations, the mean value of predictions is the final prediction and the standard deviation of predictions is the model uncertainty. 

The aleatoric uncertainty is also known as statistical uncertainty as it takes into account the uncertainty inherent in our training database and the error made by simplistic models unable to fit complex data. One strategy to capture this uncertainty is to assume that such distribution is Gaussian and dedicate one output of the neural network to estimate the variance ($\sigma_\theta^2$) 
via maximum likelihood estimation \citep{Kendall2017}:
\begin{equation}
    \mathcal{L}=1/N\sum_i^N (f_\theta(x_i)-y_i)^2/2\sigma_\theta^2 + \log\sigma_\theta^2/2.
\end{equation}
By including the variance in the formulation of our loss function, this uncertainty is learned by the neural network during the training process. Then, the total uncertainty is the square root of the sums of the epistemic and aleatoric uncertainty squared. The uncertainty can be assumed constant for every point (homoscedastic) or can vary with the input (heteroscedastic). We have used the latter assumption because the error of the neural network increases with the signal as the probability of removing a signal with higher amplitude is also higher (something that can be identified in the residuals).

We have used the recently published python-Keras package astroNN\footnote{\url{https://github.com/henrysky/astroNN}} developed by \cite{Leung2019,Mackereth2019}. To implement this new methodology we have taken our topology and modified the last layer so that it has two outputs. In our case, we have obtained better results using a new small network made of eight  convolutional (gray) blocks to estimate the uncertainty and our initial network (architecture displayed in Fig.~\ref{fig:topology}) to estimate the clean image. {In the ideal case, where the neural network is able to obtain the perfect clean version, the value of $\sigma_\theta$ will be the inherent noise of the data because the network is comparing the clean image with a noisy output. In the real case, the value of $\sigma_\theta$ will be higher than the noise, so the deviation from the known noise level will capture the error of the network, a quantity that perfectly matches with the residuals calculated between the output of the network and the original image (see Fig.~\ref{fig:uncertainty3} in the Appendix). Following \cite{Leung2019}, we have redefined the variance of the likelihood as the total of the variance inferred by the network and the intrinsic noise present in the data: $\sigma_{\theta}^2=\sigma_{net}^2+\sigma_{noise}^2$.} We want to note that the training process of this network was more difficult and slower than the previous implementation. The problem may lie in the contribution of each term of the loss function during the optimization.

Finally, in order to do the Monte Carlo, we have to insert a dropout layer for every weight layer of the network (except with the first and last layers). The free parameter, dropout rate, is  chosen to produces the expected error, which is the average difference between the result of different neural networks. We have found that a value of 0.01 usually gives good results \citep{Gal2015b}.

\begin{figure}[!ht]
\centering
\includegraphics[width=\linewidth]{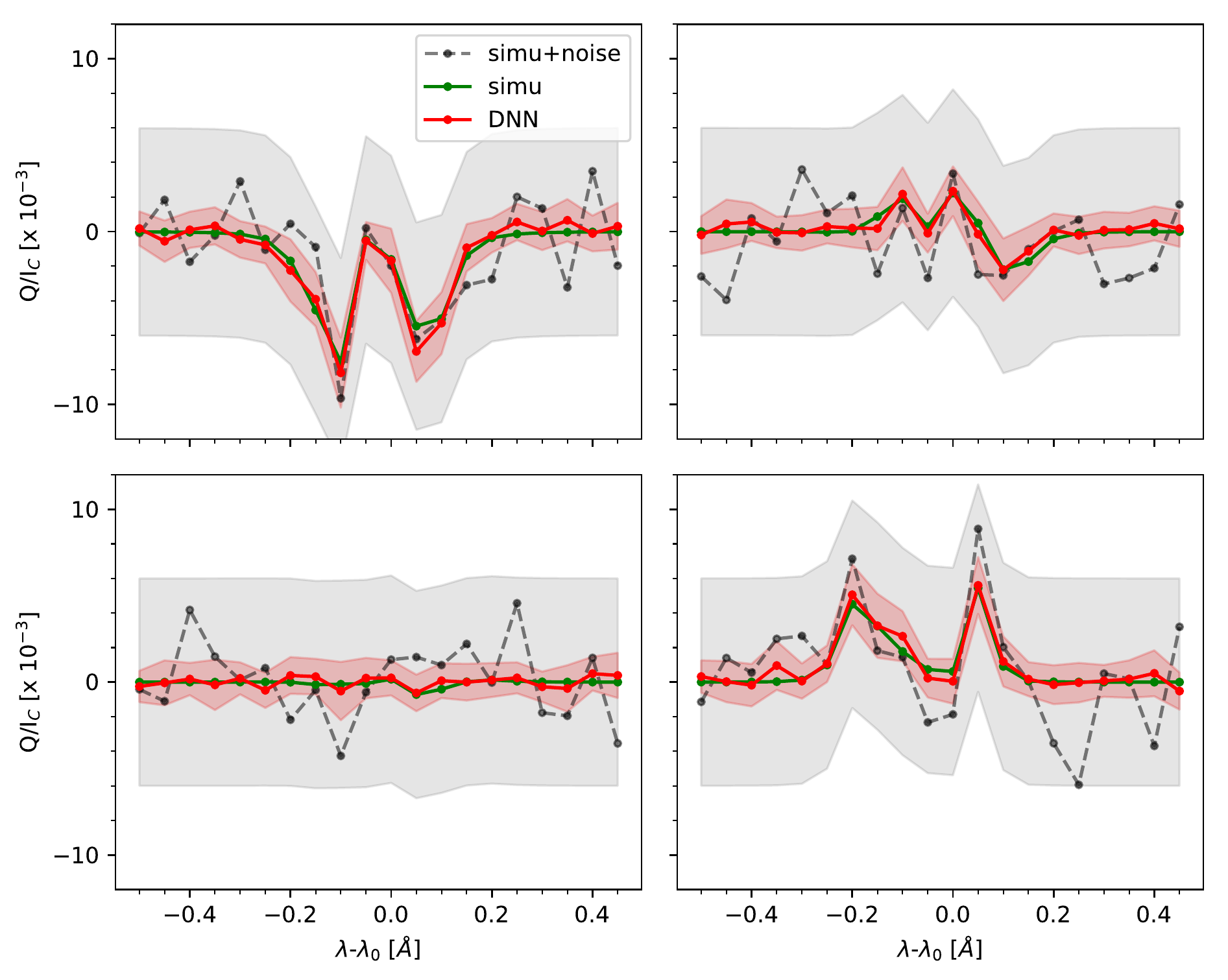}
\includegraphics[width=\linewidth]{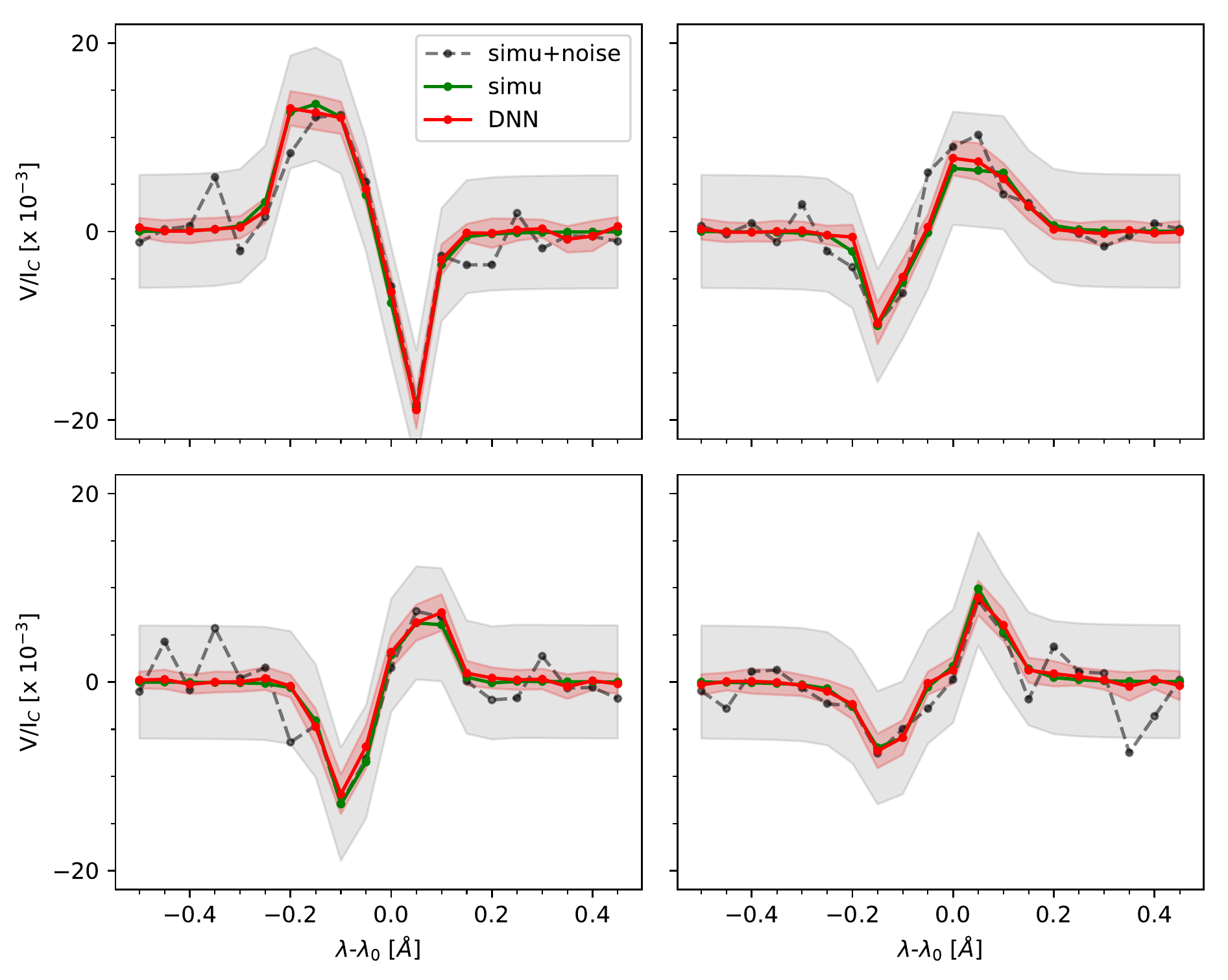}
\caption{Stokes $Q$ and $V$ profiles extracted from 4 different points of the map: the synthetic profiles (in green), with noise (in gray) and the output of the neural network (in red), are shown with their respective color bands indicating twice the uncertainty. The reference wavelength is $\lambda_0=6301.5\AA$.}
\label{fig:Sinteuncertainty}
%/scratch/carlos/DEEPL/DENOISEPAPER/final/simulation/bayes_unet/plotPerfiles.py
\end{figure}

After training this new network, we obtain clean images very similar to those shown in Fig.~\ref{fig:synthetic} and the uncertainty associated to each point (an image showing its spatial distribution is displayed in Fig.~\ref{fig:uncertainty3} in the Appendix). An example of this output is shown for a selection of four profiles in Fig.~\ref{fig:Sinteuncertainty}. We have chosen profiles with a very low signal-to-noise ratio to properly verify the efficiency of the neural network. The uncertainty returned by the network is of the order of $6\times10^{-4}I_c$, thus fulfilling that most of the simulation (green) points remain within the red shaded band estimated by the network. With this synthetic test, we have finally not only demonstrated the capabilities of the network to produce a good estimation of the clean image, but we have also shown the ability of this new neural network to estimate the uncertainty assuming Gaussian noise statistics.

\begin{figure*}[!ht]
\centering
\includegraphics[width=\linewidth]{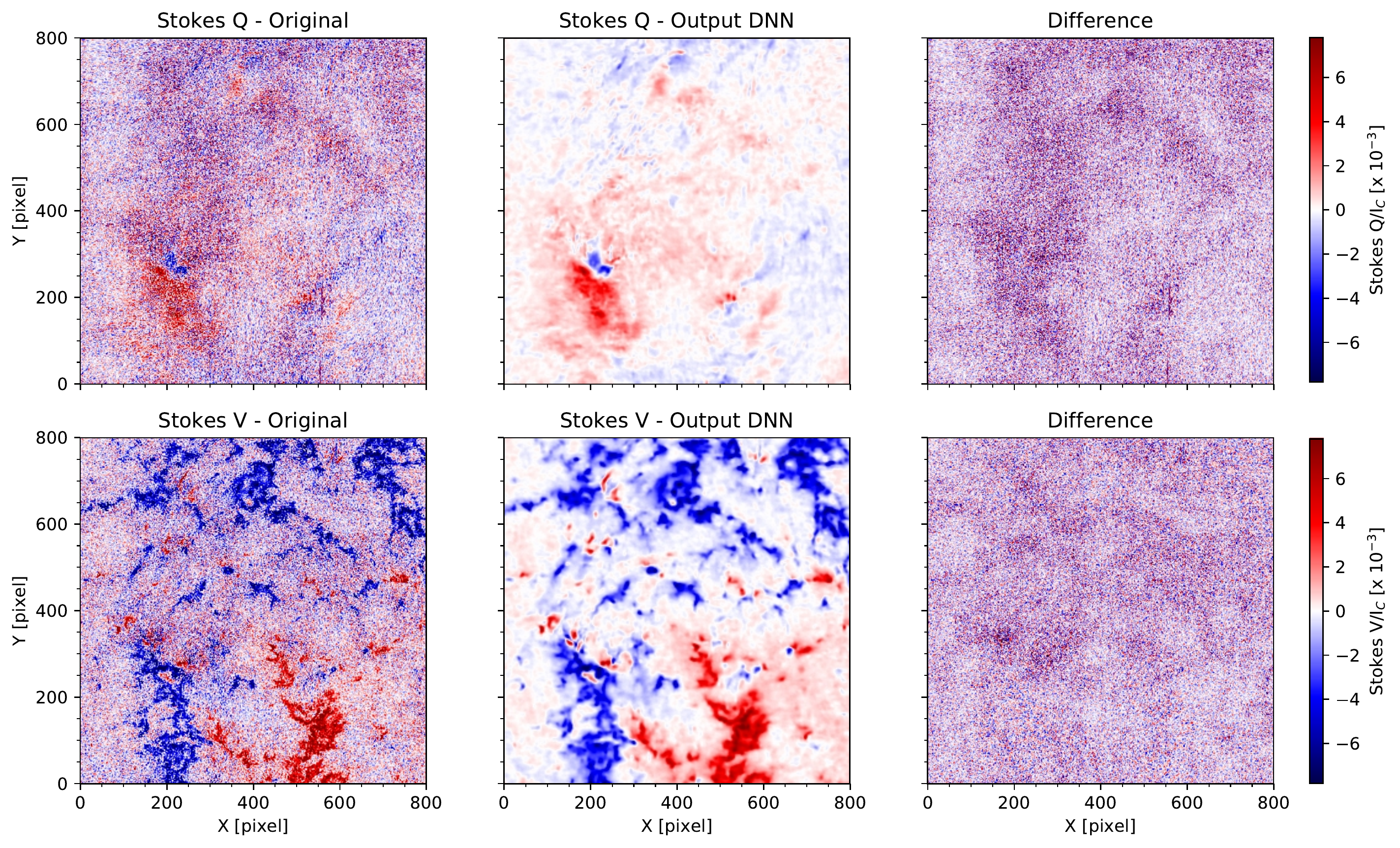}
\caption{Results for different wavelengths and Stokes parameters after apply the neural network to real observations. For Stokes $Q$ at $-0.17$\,\AA\ (top panel) and an example of Stokes $V$ (lower panel) at $-0.765$\,\AA.}
% The panels have been clipped to $\pm 8\times10^{-3}I_c.$
\label{fig:sst}
% /scratch/carlos/DEEPL/DENOISEPAPER/reales/plots_paper
\end{figure*}

%%%%%%%%%%%%%%%%%%%%%%%%%%%%%%%%%%%%%%%%%%%%%%%%%%%%%%%%%%%%%%%%%%%%%%%%%
\subsection{Real observations}\label{sec:real}
%%%%%%%%%%%%%%%%%%%%%%%%%%%%%%%%%%%%%%%%%%%%%%%%%%%%%%%%%%%%%%%%%%%%%%%%%

Perhaps the best observational test case to apply our denoising technique are Fabry-Perot observations. These two-dimensional instruments allow for the measurement of a 2D field of view in narrow-band images across the profile of the spectral line of interest. The 2D images at each wavelength are obtained successively at different times and, therefore, the spatial coherence of the signals is implicit in this technique. We will take advantage of the temporal redundancy of some observations to generate a dataset where the main difference between two time-steps is the noise distribution, as we showed in the synthetic case.

We have trained a network using two datasets observed with the CRisp Imaging SpectroPolarimeter \citep[CRISP;][]{Scharmer2006,Scharmer2008} instrument at the Swedish 1-meter Solar Telescope \citep[SST;][]{scharmer2003} in full-Stokes mode. We have used datasets that cover a variety of solar conditions from quiet Sun to more active regions. They correspond to observations taken on 2013-07-22 from 08:33 to 08:58~UT and the other one on 2013-07-19 from 13:34 to 13:59~UT. The latter dataset was also used by \citet{2017A&A...599A.133A}. These observations consist of two time series in the \ion{Ca}{ii} line at 8542\,\AA, with slightly varying seeing conditions.

The observations were reduced following the CRISPRED pipeline \citep{CRISPRED2015}, that includes: dark current subtraction, flat-field correction, and subpixel image alignment. The data were reconstructed with Multi-Object Multi-Frame Blind Deconvolution \citep[MOMFBD;][]{MFBD2002,vanNoort2005}. The MOMFBD code applies a Fourier filter to the reconstructed images that suppresses frequencies above the diffraction limit of the telescope and which is slightly modified according to each patch of the image. This processing will generate an additional correlation between the signal and the noise of each region.

We use these images as the input and output of our training set. A total of N = 10000 patches of 52$\times$52 pixels are randomly extracted from the temporal series. They are also randomly extracted from the spectral positions and from different Stokes parameter ($Q$, $U$ and $V$). We also randomly extracted a smaller subset of 1000 patches which will act as a validation set to check that the CNN generalizes well. {Again, the network is trained during 20 epochs.}

A crucial ingredient for the success of this process is the generation of a suitable training set of high quality. Physical data augmentation has been important in reducing the possible effects of the quick evolution of signals between two time-steps. For that, we apply rotations, sign changes, etc. In this way, we can, not only generate a dataset with much more variety but generate opposite states (for example a region where the signal increases can be used in reverse order). With that, we achieve that the average behavior remains constant and the noise is the only different factor between two frames. In the case of a quick evolution of the signals between two frames, we might take into account the motion (e.g. calculating the optical flow) and wrap one frame to the other \citep{Ehret2018}, however it does not seem to be really necessary because the training data have been chosen stable over time.

%%%%%%%%%%%%%%%%%%%%%%%%%%%%%%%%%%%%%%%%%%%%%%%%%%%%%%%%%%%%%%%%%%%%%%%%%
\subsubsection{Monochromatic maps}\label{sec:2dmaps}
%%%%%%%%%%%%%%%%%%%%%%%%%%%%%%%%%%%%%%%%%%%%%%%%%%%%%%%%%%%%%%%%%%%%%%%%%

Once the network was trained it was used on observations of a region of flux emergence taken on 2016-09-21 from 12:38 to 12:58~UT\footnote{An example of the neural network applied to the same observations used for the training is shown in the Appendix \ref{fig:validation}}. Figure~\ref{fig:sst} displays the map estimated by the network for different wavelengths and Stokes parameters. For Stokes $Q$ (top panel) we have chosen the wavelength at the core of the spectral line due to its weak signal. In the case of Stokes $V$ (lower panel), we have chosen a wavelength closer to the wing where the noise effect is more evident as Stokes $V$ has usually higher signals.

Like in the synthetic case, we have examined that we do not lose signals during reconstruction (see the right column of Fig.~\ref{fig:sst}). Although the amplitudes of the noise pattern correlates with the strength of the signals, the local spatial average is zero. This result has not been imposed on the training but is successfully obtained at the end of the training.

In this unsupervised training, the network is also not only able to infer the noise from the image but to also eliminate artifacts generated for example by the post-processing. In the case of Stokes $Q$, vertical features appear in the pixels with x=580 and are visible in the original image but not in the reconstruction of the neural network. The neural network is able to solve this task because although some artifacts are located in the same location as the sensor, the images used for training have been slightly rotated to compensate for the solar rotation.

\begin{figure}[!ht]
\centering
\includegraphics[width=\linewidth]{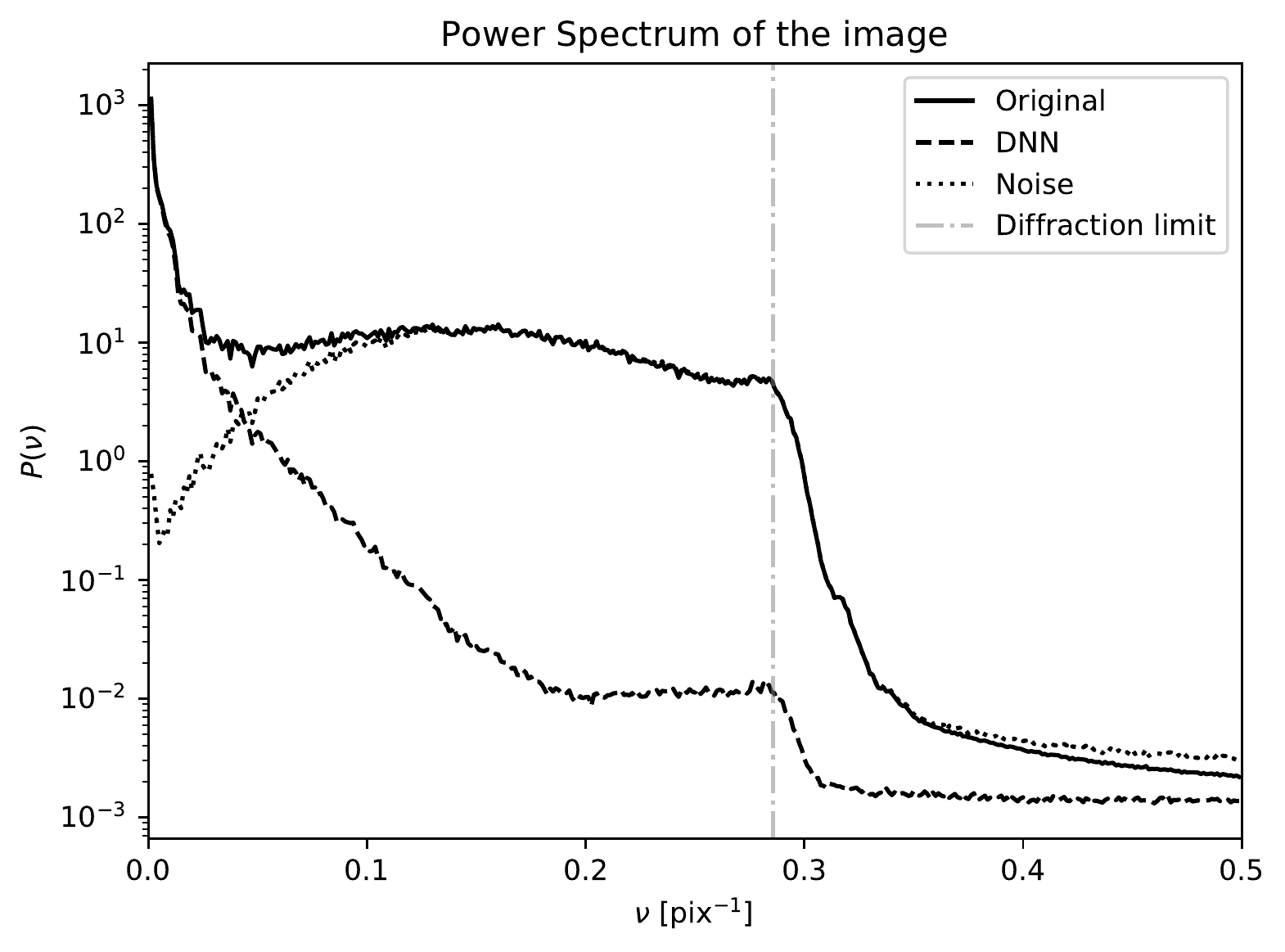}
\caption{{Spatial Fourier power spectra of the original image, the output of the neural network and the difference between them, which is our estimate of the noise. We have marked the diffraction limit of the telescope as a reference.}}\label{fig:power}
%/scratch/carlos/DEEPL/DENOISEPAPER/final/reales/powerplot/FTSdatos.py
\end{figure}

Another interesting point to study is the range of spatial scales at which the neural network is performing. Figure~\ref{fig:power} displays the power spectrum of the images shown in the first row of Fig.~\ref{fig:sst}. For the second row, the result is very similar. The difference between the power spectrum of the original image and the output starts around $\nu$ = 0.05\,pix$^{-1}$, which means that the neural network is operating mainly at scales smaller than 20 pixels, where the noise is present. As the noise is correlated with the signals, it does not have a flat spectrum, but a complex one as shown with a dotted line in Fig.~\ref{fig:power}. The spectrum of the clean image decreases down to $\nu$ = 0.2\,pix$^{-1}$ where it flattens and then falls suddenly after the diffraction limit. This indicates that the network keeps and enhances structures as small as 4 or 5 pixels.

%%%%%%%%%%%%%%%%%%%%%%%%%%%%%%%%%%%%%%%%%%%%%%%%%%%%%%%%%%%%%%%%%%%%%%%%
\subsubsection{Stokes profiles}\label{sec:profiles}
\begin{figure}[!ht]
\centering
\includegraphics[width=\linewidth]{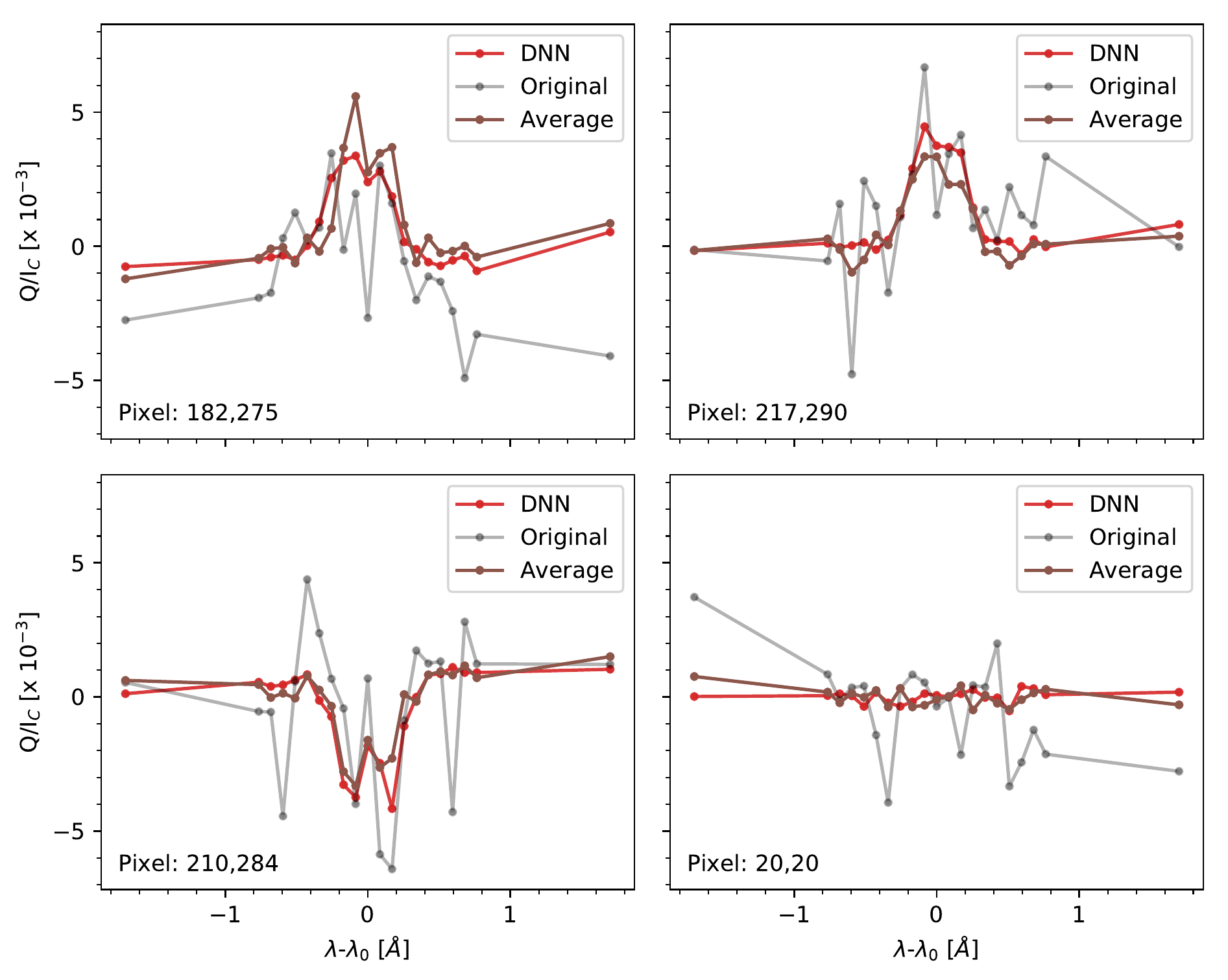}
\caption{Stokes $Q$ profiles extracted from 4 different points of the map: the original (in gray), the reconstructed by the neural network (in red) and the average profile in those pixels of the entire time series (in brown). The reference wavelength is $\lambda_0=8542.1\AA$.}
\label{fig:minimosaico}
%/scratch/carlos/DEEPL/DENOISEPAPER/reales/miniMosaico
\end{figure}

%%%%%%%%%%%%%%%%%%%%%%%%%%%%%%%%%%%%%%%%%%%%%%%%%%%%%%%%%%%%%%%%%%%%%%%%%
Once all wavelength images were cleaned, we studied the spectral profile at each point on the map. Figure~\ref{fig:minimosaico} shows the spectral profile of some points of the FOV before and after reconstruction. The most surprising aspect of using this technique is the high spectral coherence obtained even without showing such information during the training. This information is implicitly contained in the original data and only with the spatial reconstruction we are able to reveal it. This spectral coherence is shown for example in Stokes $Q$ in the form of a symmetric profile where the wings tend to zero, something that was never imposed. This may seem trivial but such behavior is not always observed in the original data given the high noise level.

We have also drawn the profile at these pixels averaged over the whole time series. The comparison shows that the reconstructed profiles are very similar to the average profiles.This indicates that we are able to reproduce a signal similar to the general behavior without losing the temporal evolution of the analyzed time frame.

The remarkable spectral coherence achieved in the $Q$ and $U$ profiles raises the question of whether we are somehow introducing it or it is already present in the data. A simple way to check this is to train the neural network using only Stokes $V$, and as expected, the result is the same. On the contrary, if the training is done with only Stokes $Q$ and $U$, the signals obtained by the network are slightly worse. Having a dataset with strong signals, at least 2 or 3 times above the noise, is critical because it makes the training more robust. The network can then differentiate  more easily when it is noise and when not. To generate a balanced training set, one can either set a threshold or take the wavelengths of the core of the spectral line to promote patches with higher signals.

\begin{figure}[!ht]
\centering
\includegraphics[width=\linewidth]{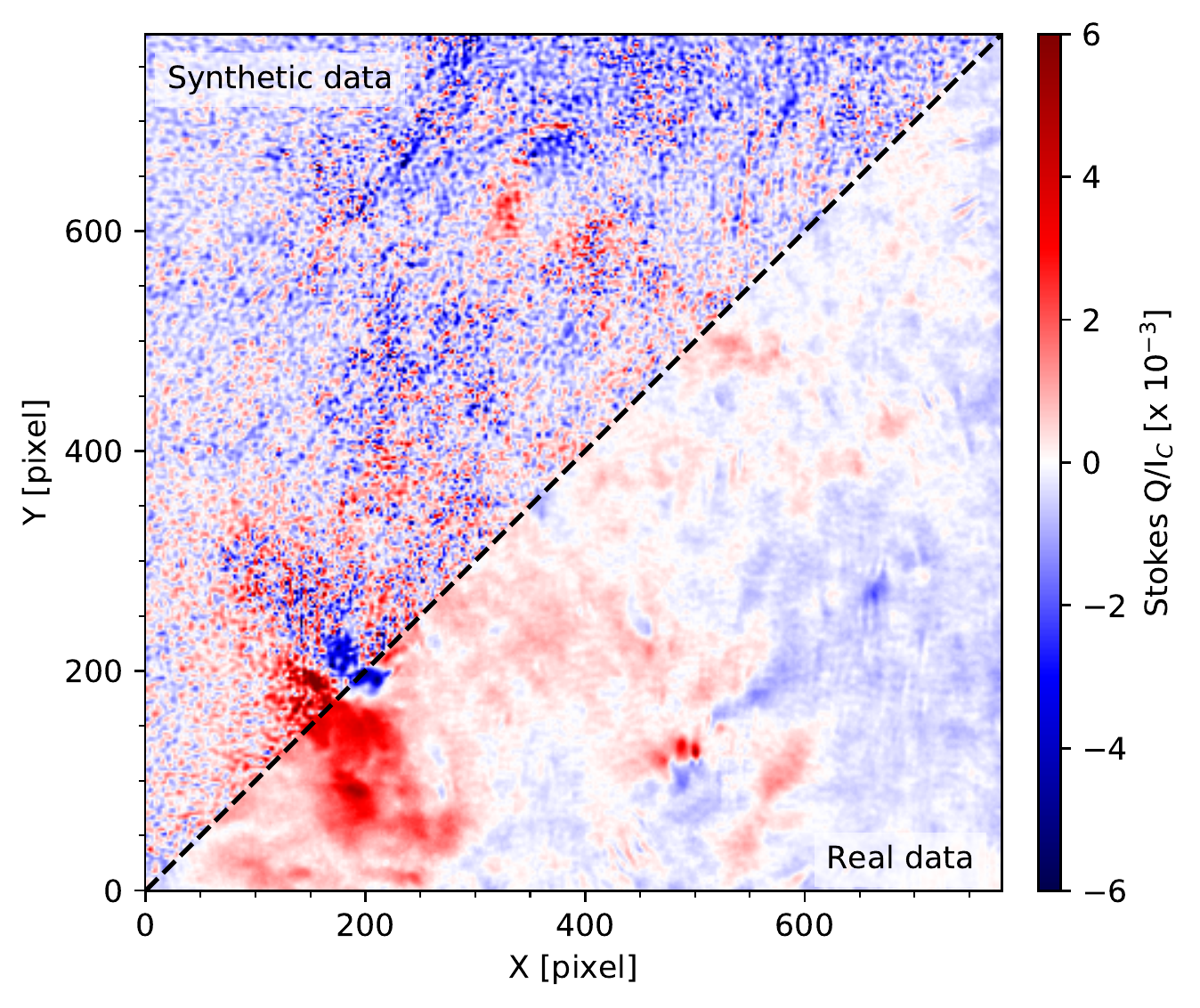}
\caption{Output of the network trained with synthetic data and Gaussian noise independent at each pixel (upper half) and with CRISP data (lower half).}
\label{fig:Halfs}
%/scratch/carlos/DEEPL/DENOISEPAPER/final/reales/simvsreal/plotDiferenciaCRISP.py

\end{figure}

Figure~\ref{fig:Halfs} illustrates how the performance of the denoising method worsens when the network is based/trained on images with a different/wrong spatial scale of noise. Here we trained the network with the synthetic data from the MHD simulation to which we added the Gaussian noise of our CRISP dataset. Since the spatial distribution of the noise in the CRISP data is very different, more non-Gaussian, only a partial recovery is achieved when such network is applied to the CRISP data.

% /scratch/carlos/DEEPL/DENOISEPAPER/reales/plots_paper/ComparaIguales.py 
% /scratch/carlos/DEEPL/DENOISEPAPER/reales/unet_sst_ONLYOBS/fromSYNTHETIC

%%%%%%%%%%%%%%%%%%%%%%%%%%%%%%%%%%%%%%%%%%%%%%%%%%%%%%%%%%%%%%%%%%%%%%%%%
% \subsubsection{Uncertainty in the inference}\label{sec:uncertainty2}
%%%%%%%%%%%%%%%%%%%%%%%%%%%%%%%%%%%%%%%%%%%%%%%%%%%%%%%%%%%%%%%%%%%%%%%%%

The Bayesian neural network implemented with the real data, reveals that the network uncertainties is of the same order of magnitude as in the synthetic case, with a typical value around $6\times10^{-4}I_c$. While the intrinsic noise of the observations is lower than in the synthetic case, the uncertainty is higher than the expected for this noise level. This is probably because of its complex level of corruption, originated probably in the MOMFBD reconstruction.

\begin{figure}[!ht]
\centering
\includegraphics[width=\linewidth]{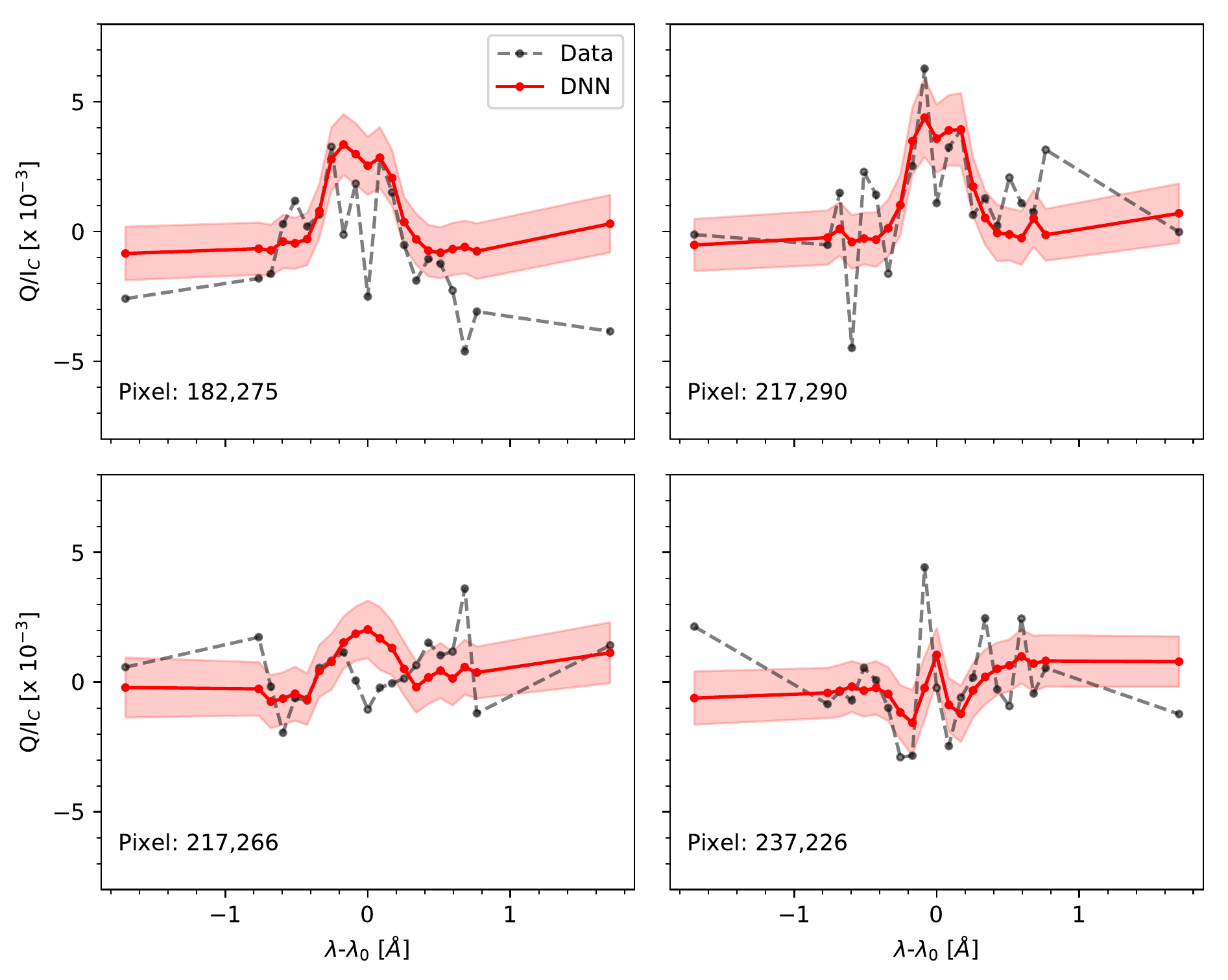}
\includegraphics[width=\linewidth]{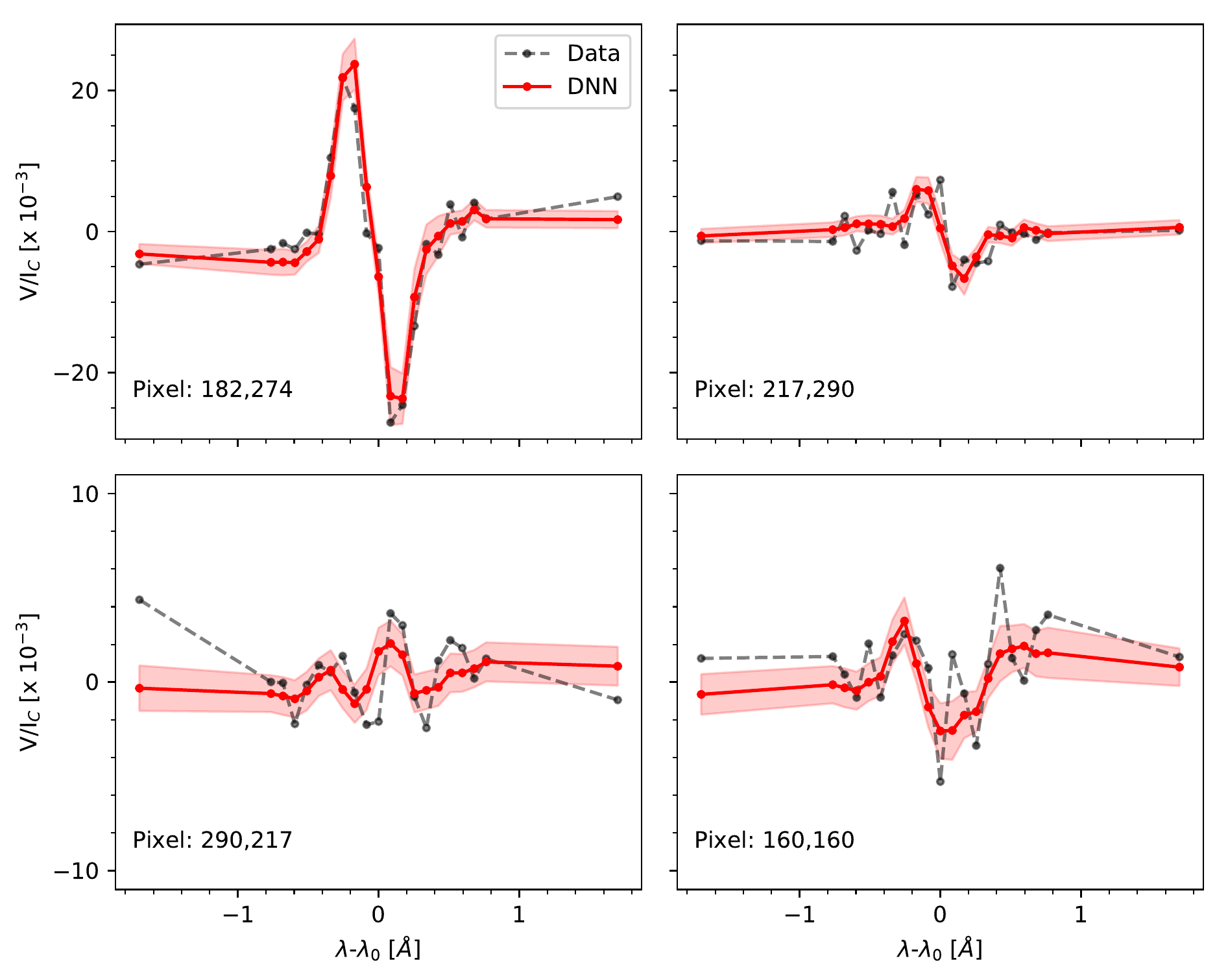}
\caption{Stokes $Q$ and $V$ profiles extracted from 4 different points of the map: the observed profiles (in gray) and the output of the neural network (in red) are shown with their respective color bands indicating twice the uncertainty calculated by the neural network.}
\label{fig:crispUncertainty}
%/scratch/carlos/DEEPL/DENOISEPAPER/sintetico2/test/plotProfilesCRISP.py
\end{figure}

An example of the prediction with the uncertainty is displayed in Fig.~\ref{fig:crispUncertainty}. This estimation is restricted to Gaussian aleatoric uncertainties. However, in reality, our noise (after the post-processing) does not fully follow a Gaussian distribution. Another problem is separating pure noise from the effects coming from evolution. So we have calculated the noise as a moving (local) standard deviation to take into account the variation of the noise with the signal. As a result this uncertainty estimation may be lower than the real one and only with the development of new techniques we will be able to have better estimations in the future \citep{Tagasovska2018}.

%%%%%%%%%%%%%%%%%%%%%%%%%%%%%%%%%%%%%%%%%%%%%%%%%%%%%%%%%%%%%%%%%%%%%%%%%
\subsubsection{Constraining the solution}\label{sec:constrain}
%%%%%%%%%%%%%%%%%%%%%%%%%%%%%%%%%%%%%%%%%%%%%%%%%%%%%%%%%%%%%%%%%%%%%%%%%

The advantage of the neural network model presented in these sections is that the inference is independent of the given wavelength or whether the image displays linear or circular polarization. This allows for a more simple and diverse training because we can use any spectral information from the observations regardless of which Stokes parameter it belongs, to generate a large training set.

One can also take into account the spectral information of the profile to improve the estimation of signals. To study this possibility, we allow the network to  not only be able to understand the spatial information but also to extract the correlation between the wavelength points and thus estimate the spectral profiles of the entire FOV at the same time. {For this purpose, we train the first neural network (the one described in  Fig.~\ref{fig:topology} without uncertainties)}, but now by using a cube of dimensions $nx\times ny\times n\lambda$ (patches of size $52\times52\times21$ pixels in this case) as our input/output. By introducing this new dimension as a network input, it gets more difficult to generate a rich variety of profiles using the same dataset. So we increased the diversity by generating profiles with the wavelength points in reverse order, thus doubling the number of existing profiles.

\begin{figure}[!ht]
\centering
\includegraphics[width=\linewidth]{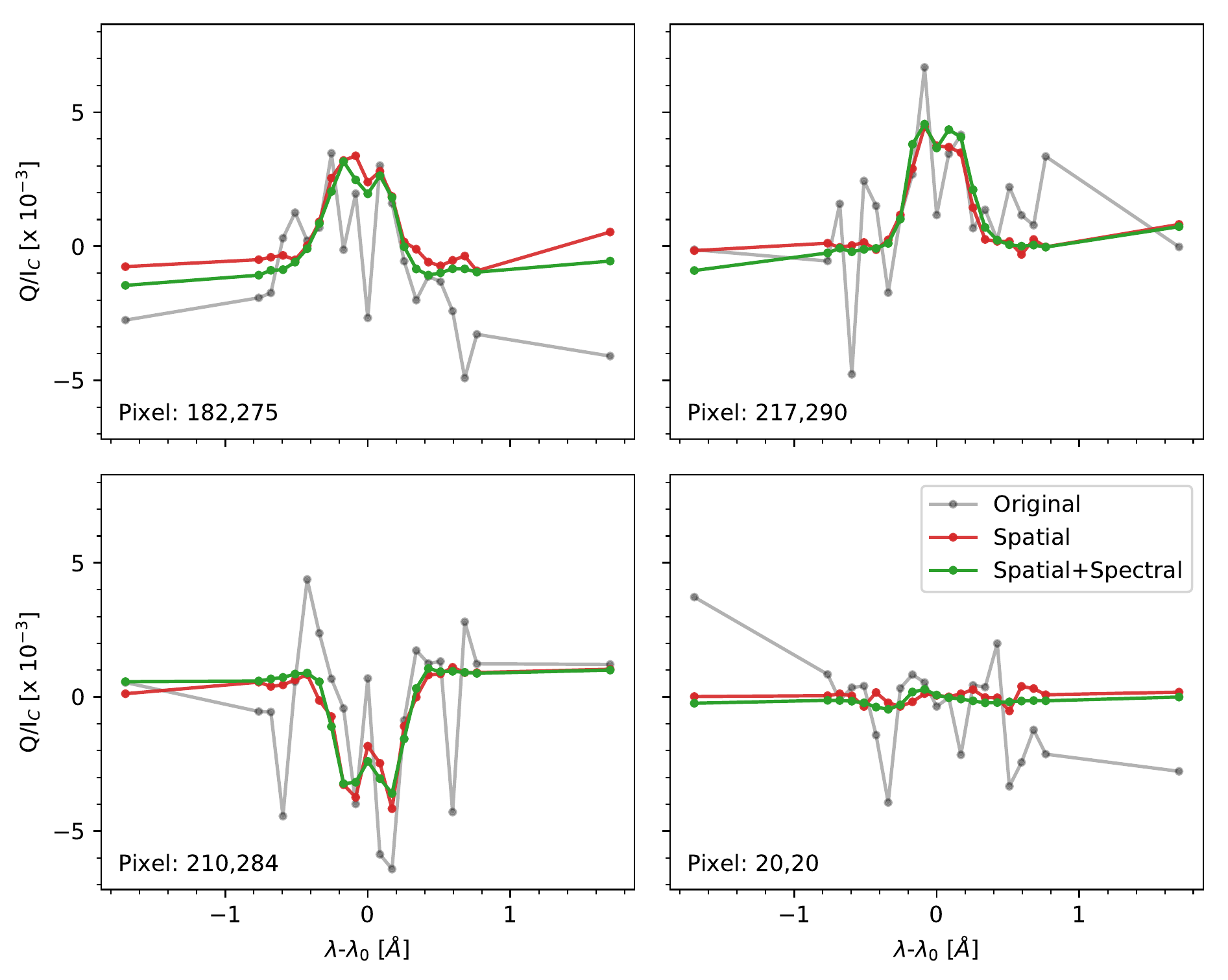}
\caption{Same Stokes profiles of Fig.~\ref{fig:minimosaico} but showing the comparison between the inference of the network trained with images (red curve) and with Stokes profiles (green curve).}
\label{fig:minimosaico2}
%/scratch/carlos/DEEPL/DENOISEPAPER/final/reales/unet_sstSPECTRAL
\end{figure}

\begin{figure}[!ht]
\centering
\includegraphics[width=\linewidth]{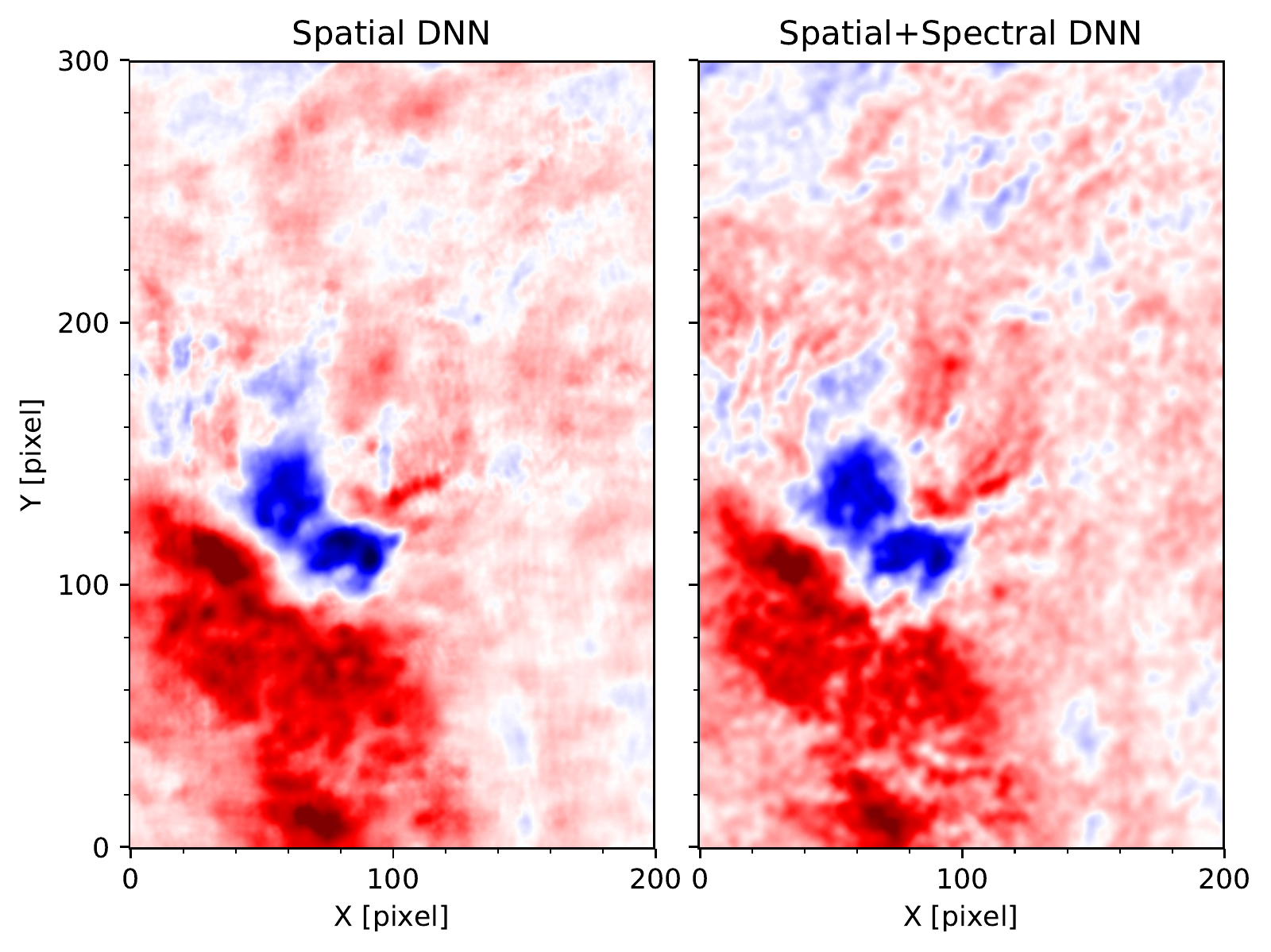}
\caption{Monochromatic images at the core of Stokes $Q$ after apply the network trained with monochromatic images (left panel) and with spectral profiles (right panel). The signals has been clipped to $\pm 8\times10^{-3}I_c$ for both panels.}
\label{fig:duo}
%/scratch/carlos/DEEPL/DENOISEPAPER/reales/unet_sst2...
\end{figure}

Figure \ref{fig:minimosaico2} shows an example of the previously chosen profiles and their inference with both methods. The new reconstruction (in green), although not very different from the previous network (in red), shows smother profiles (even when the signal is as small as the last profile). This improvement is also detected in the two-dimensional maps. Figure~\ref{fig:duo} shows the lower right region of the FOV reconstructed with both methods: only spatial (left panel) and adding spectral information (right panel). The average difference between them is around $1.10^{-4}I_c$. The right panel appears with a better-defined structure. An example of this is an intrusion of a penumbral filament into the umbra and a lot more structure visible in the background.

The neural network is now specific to the spectral line used and its sampling (distance and number of wavelength points). This neural network would no longer be general given that now the network encodes specific spectral information and we would need a different one for a different observational configuration or spectral line. Also, we are not certain if the neural network treats the spectral information in the correct way. It might correlate wavelength points which are far apart and not really physically connected. Also, the training set that was chosen to be stable, might not contain enough information to recognize cases with strong velocities.

Lastly, another information that we could take advantage of would be the availability of Stokes parameters. This may be a good idea but the correlation between the Stokes parameters is very low. Some remaining correlation may exist due to the demodulation process or some physical process that acts in some cases \citep[e.g. the alignment to orientation conversion;][]{Kemp1984}, but in general the magnetic field will produce different signals given its geometry and the parameters will be independent.

We have tested this idea by training a network that takes as an input all Stokes parameters $Q$, $U$, and $V$. The training of this network is more difficult because the network must combine the information of $\sim$60 points in wavelength and spatially to generate a good inference of another $\sim$60 points. Unfortunately, the result is not as good as in previous tests. In this case the residuals appear patchy with amplitudes of the order of magnitudes of the original signal. We probably need a larger network, a more diverse set of training or a more robust training mode, but with these preliminary results, we have not seen the need for further progress using all parameters at the same time.

%%%%%%%%%%%%%%%%%%%%%%%%%%%%%%%%%%%%%%%%%%%%%%%%%%%%%%%%%%%%%%%%%%%%%%%%%
\subsection{Other instrumentation}\label{sec:gregor}
%%%%%%%%%%%%%%%%%%%%%%%%%%%%%%%%%%%%%%%%%%%%%%%%%%%%%%%%%%%%%%%%%%%%%%%%%

This technique could also be applied to slit instruments. Slit-spectropolarimeters provide spectra with all the wavelengths obtained simultaneously, but, only one slice of the field of view at any given time. This means that observing a 2D region of the solar surface requires the construction of a raster map by shifting the slit in small increments perpendicular to its direction. In this case, the retrieved 2D map does not correspond to a single moment, but its measurement spans during a certain temporal range. For slow evolving processes, the whole map has a spatial coherence that can also be exploited by this technique. However, Slit-spectropolarimeters provide spectra with a much better spectral resolution and a larger spectral range. This implies that the network will take more time to clean the spectra given the number of wavelength points.

If we repeat the previous exercise of training a neural network with the full spectral range, it will be a very complicated process given the number of points if the same topology is used. Therefore, for slit-instruments, we would recommend using more robust algorithms such as PCA, which use the spectral information to reduce the noise and without the mentioned limitations.

%%%%%%%%%%%%%%%%%%%%%%%%%%%%%%%%%%%%%%%%%%%%%%%%%%%%%%%%%%%%%%%%%%%%
\section{Discussion and perspectives}
%%%%%%%%%%%%%%%%%%%%%%%%%%%%%%%%%%%%%%%%%%%%%%%%%%%%%%%%%%%%%%%%%%%%%%%%%
In this study, we have proposed a new approach for recovering signals under a complex noise corruption (photon noise, instrumental and post-processing artifacts) using neural networks. The training of the network is done without a priori knowledge of clean signals or the corruption level, but only using the same observations as the generative model. This approach yields results with the similar quality as compared with the case when the clean target is given. We have shown that this technique allows us to reveal signals that were strongly affected by the noise, and we have shown some examples of the improvement of typical signals obtained in current telescopes, such as the Swedish 1-m Solar Telescope. Although we have used this telescope as an example, given the generality of the technique, it can be applied to any telescope, instrument, solar region or physical quantity.

The results presented in the present study are particularly relevant for detecting weaker signals and thus studying the magnetization of regions in the chromosphere and corona. Although the network architecture that we used is very simple, it represents the first step toward a suite of new tools to remove artifacts, fringes, and denoise observational data. Our method intrinsically makes use of the sparsity present in solar data \citep{Sparse2015}, to separate signals from the noise in the spatial dimension. We have shown both, a method that uses only spatial information and another one in which the spectral information is also added. We have concluded, however, that PCA techniques seem like a better option to decrease noise along the spectral direction.

Despite the successful neural networks presented here, we anticipate that improvements in the quality of the reconstruction can be achieved:
\begin{enumerate}
\item there is still one dimension that we have not exploited and that contains information that can be used to further reduce noise: using the temporal coherence during the prediction. In fact, this last idea is being exploited in videos to make superresolution \citep{TempoGAN2018}.
\item another merit function could be used to reduce the spatial blurring in the predictions of the network, for example by using another neural network as the discriminator as it is done in Generative Adversarial Networks \citep[GANs;][]{Ledig2016}. There are more complex networks, such as the MSD-net architecture which uses dilated convolutions to learn features at different scales. This approach has even better performance than the U-net model with small training sets \citep{Pelt2018}. 
\item it is also fundamentally important to improve the uncertainty estimation by calibrating the network with new methods that are valid for non-Gaussian noise distributions \citep{Maddox2019}. 
\end{enumerate}
Finally, including a quick implementation of the optical flow to counteract the changes between two frames could help to better preserve the signals and avoid them being blurred by the network \citep{Ehret2018}.

\begin{acknowledgements}
{We would like to thank the anonymous referee for the comments and suggestions.}
This research has made use of NASA's Astrophysics Data System Bibliographic Services.
CJDB thanks Andr\'es Asensio Ramos, Carolina Robustini and Flavio Calvo for initial discussions about this project. 
JdlCR is supported by grants from the Swedish Research Council (2015-03994), the Swedish National Space Board (128/15). 
This project has received funding from the European Research Council (ERC) under the European Union's Horizon 2020 research and innovation programme (SUNMAG, grant agreement 759548). 
SD and JdlCR are supported by a grant from the Swedish Civil Contingencies Agency (MSB). SD acknowledges the Vinnova support through the grant MSCA 796805.
The Institute for Solar Physics is supported by a grant for research infrastructures of national importance from the Swedish Research Council (registration number 2017-00625).
%SST
The Swedish 1-m Solar Telescope is operated on the island of La Palma by the Institute for Solar Physics of Stockholm University in the Spanish Observatorio del Roque de los Muchachos of the Instituto de Astrofísica de Canarias. This research has made use of NASA’s Astrophysics Data System Bibliographic Services.
We acknowledge the community effort devoted to the development of the following open-source packages that were used in this work: numpy (\url{numpy.org}), matplotlib (\url{matplotlib.org}) and keras (\url{keras.io}).
\end{acknowledgements}

% % \clearpage
% \bibliographystyle{aa}
% \bibliography{../general}

% \end{document}


% \clearpage
% \newpage

% \clearpage
\bibliographystyle{aa}
% \bibliography{general}

\begin{appendix}

\section{Uncertainty calibration}

\begin{figure}[!ht]
\centering
\includegraphics[width=\linewidth]{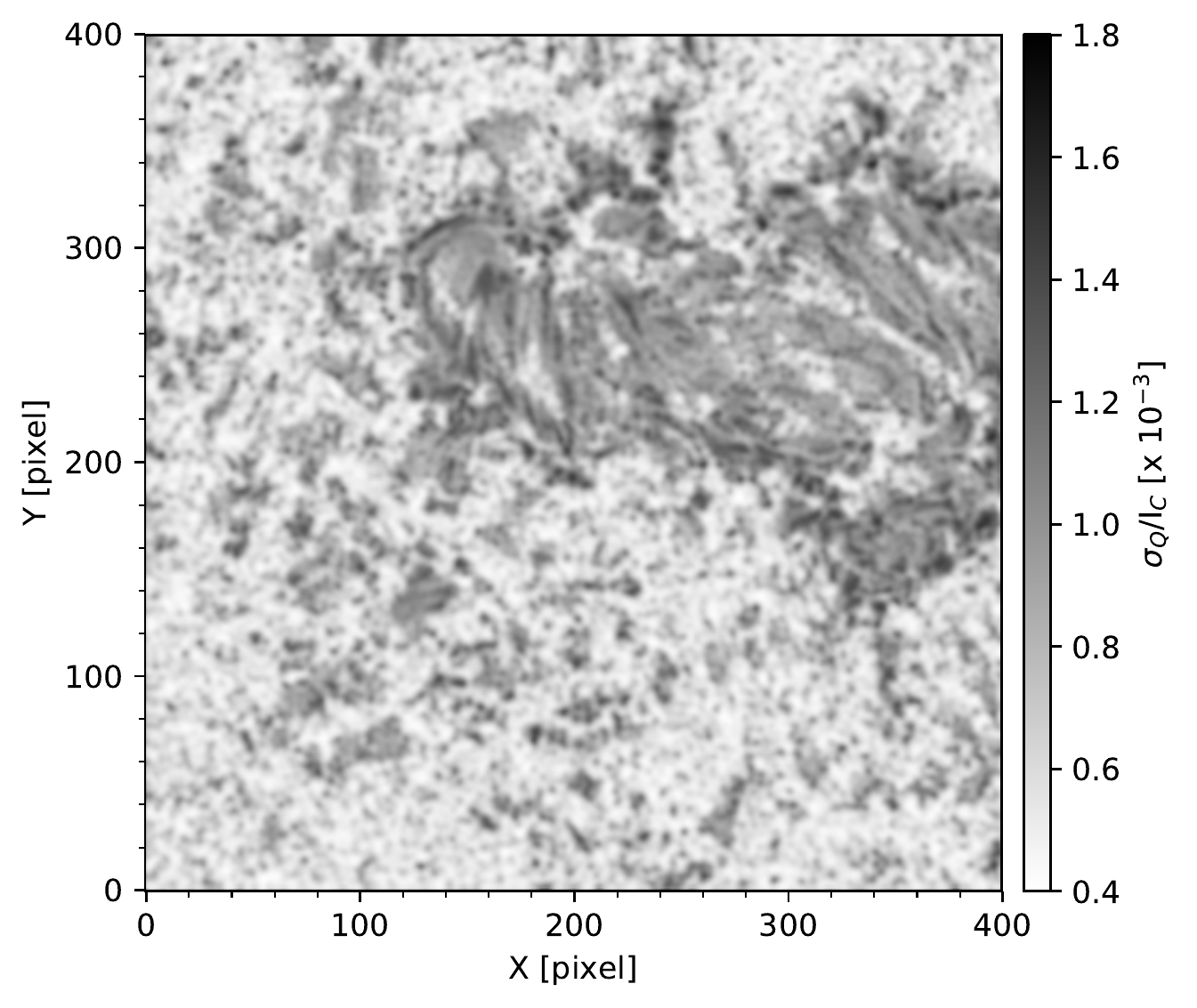}
\includegraphics[width=\linewidth]{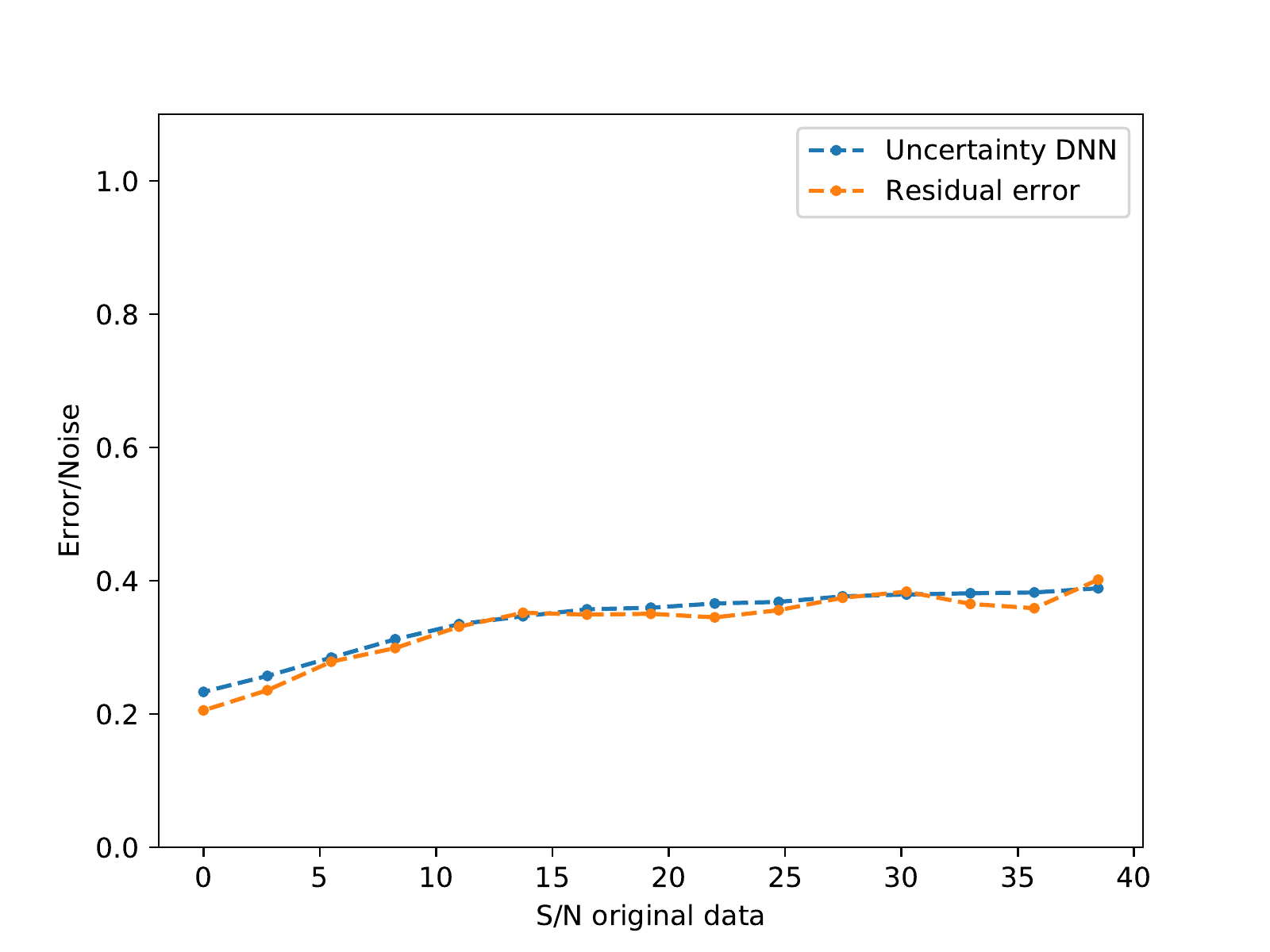}
\caption{Spatial distribution of the uncertainty calculated by the neural network (upper panel) and the comparison with the residuals calculated with the original image as a function of the amplitude of the signal (lower panel).}
\label{fig:uncertainty3}
%/scratch/carlos/DEEPL/DENOISEPAPER/reales/unet_sst2/plotNOISEpaper.py
\end{figure}

At this point, we show the spatial distribution of the uncertainty estimated by the network in Fig~\ref{fig:uncertainty3}. To demonstrate that the uncertainty is well calibrated, we display in the lower panel also the residuals calculated as the difference between the original image and the one inferred by the neural network. This comparison provides the confidence to know that our network is inferring a good estimation of the uncertainty.

\section{Network architecture}
{The neural network has been developed using the Keras Python library, with the Tensorflow backend for the computations. A detailed description is given in the Table\,\ref{tab:arquitecture}. For the image-image training we have used $n=32$ and $m=1$. For the \ion{Ca}{ii} full profile training we have used $n=32$ and $m=21$. In the case one wants to calculate the uncertainty, we need to use the first topology for the inference and the one described in Table\,\ref{tab:arquitecture2} for the variance inside the same network, so the network has one input but two different outputs.}

\begin{table}[!ht]
\begin{tabular}{ccc}
Layer & \# filters & Function \\
\hline\hline
Conv2D+ReLU 1 & $n$ & Conv 3$\times$3 + ReLU \\
Conv2D+ReLU 2 & $n$ & Conv 3$\times$3 + ReLU\\
MaxPooling 1 & $n$ & Maxpooling 2$\times$2 \\
Conv2D+ReLU 3 & $n\times2$ & Conv 3$\times$3 + ReLU\\
Conv2D+ReLU 4 & $n\times2$ & Conv 3$\times$3 + ReLU\\
MaxPooling 2 & $n\times2$ & MaxPooling 2$\times$2\\
Conv2D+ReLU 5 & $n\times4$ & Conv 3$\times$3 + ReLU\\
Conv2D+ReLU 6 & $n\times4$ & Conv 3$\times$3 + ReLU\\
Upsampling 1 & $n\times4$ & Upsampling 2$\times$2\\
Conv2D+ReLU 7 & $n\times2$ & Conv 3$\times$3 + ReLU\\
Concatenate 1 & $n\times2$ & Concat with Conv2D+ReLU 4 \\
Conv2D+ReLU 8 & $n\times2$ & Conv 3$\times$3 + ReLU\\
Conv2D+ReLU 9 & $n\times2$ & Conv 3$\times$3 + ReLU\\
Upsampling 2 & $n\times2$ & Upsampling 2$\times$2\\
Conv2D+ReLU 10 & $n$ & Conv 3$\times$3 + ReLU\\
Concatenate 1 & $n$ & Concat with Conv2D+ReLU 2 \\
Conv2D+ReLU 11 & $n$ & Conv 3$\times$3 + ReLU\\
Conv2D+ReLU 12 & $n$ & Conv 3$\times$3 + ReLU\\
Network Output & $m$ & Conv 1$\times$1 \\
\end{tabular}
\caption{Network architecture used in our experiments. Number of network input feature maps $n$ and the  dimension of the output $m$  depend on the experiment.}
\label{tab:arquitecture}
\end{table}

\begin{table}[!ht]
\begin{tabular}{ccc}
Layer & \# filters & Function \\
\hline\hline
Conv2D+ReLU 1 & $n$ & Conv 3$\times$3 + ReLU \\
Conv2D+ReLU 2 & $n$ & Conv 3$\times$3 + ReLU\\

Conv2D+ReLU 3 & $n$ & Conv 3$\times$3 + ReLU\\
Conv2D+ReLU 4 & $n$ & Conv 3$\times$3 + ReLU\\

Conv2D+ReLU 5 & $n$ & Conv 3$\times$3 + ReLU\\
Conv2D+ReLU 6 & $n$ & Conv 3$\times$3 + ReLU\\

Conv2D+ReLU 7 & $n$ & Conv 3$\times$3 + ReLU\\
Conv2D+ReLU 8 & $n$ & Conv 3$\times$3 + ReLU\\

Conv2D 13 & 1 & Conv 1$\times$1 \\

Variance Output & - & $-$Abs(x)$-12$\\
\end{tabular}
\caption{Network architecture to calculate the variance. Given that the variance in log units is a big negative number, to accelerate the convergence of the network we had to use an additional layer to constrain its value.}
\label{tab:arquitecture2}
\end{table}

\section{Validation set}
Here we show some maps similar to the one presented in previous sections where the neural network is applied to the same observations used for the training process. In this map, we can verify the generalization during the training process.

\begin{figure*}[!ht]
\centering
\includegraphics[width=0.9\linewidth]{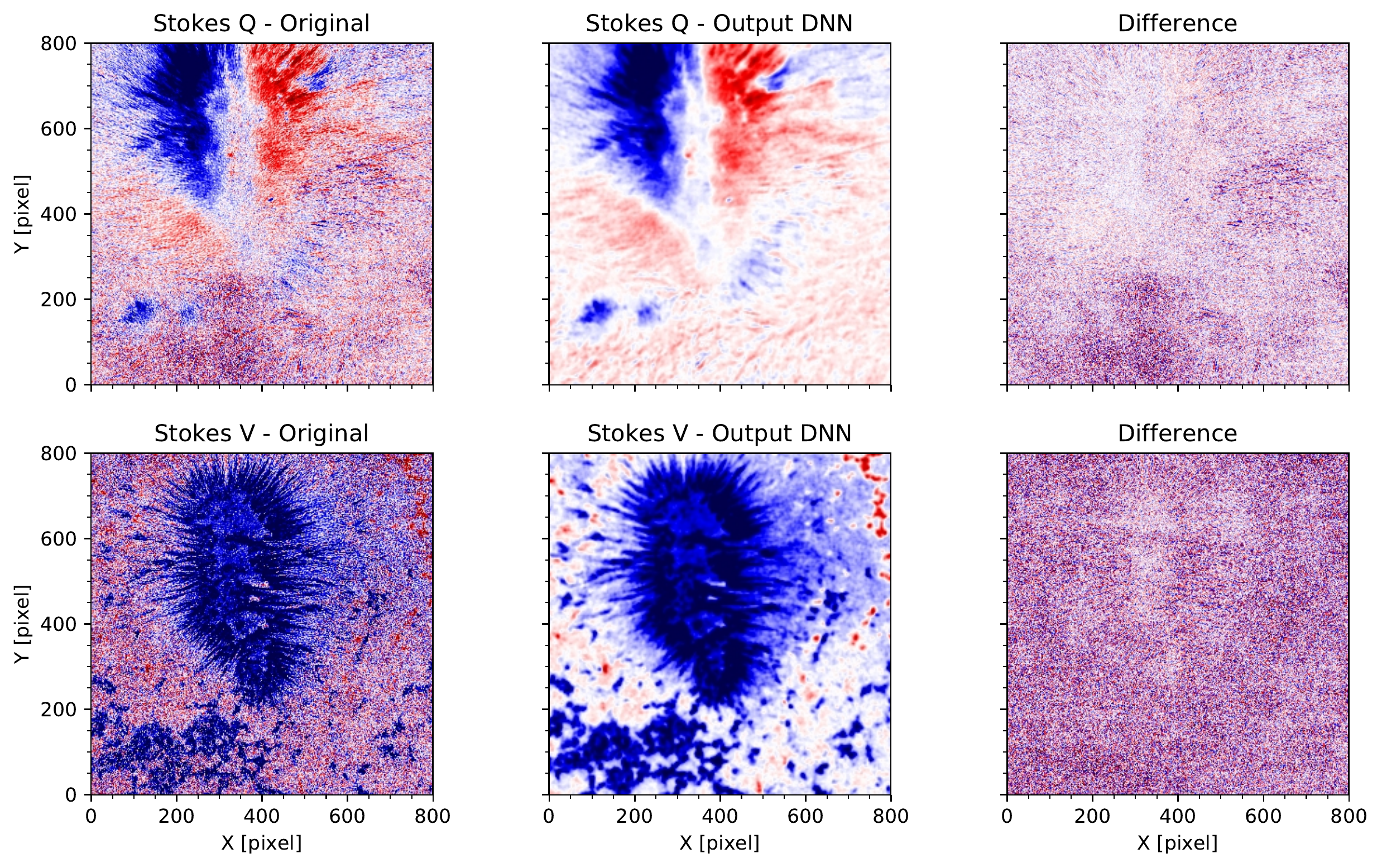}
\includegraphics[width=0.9\linewidth]{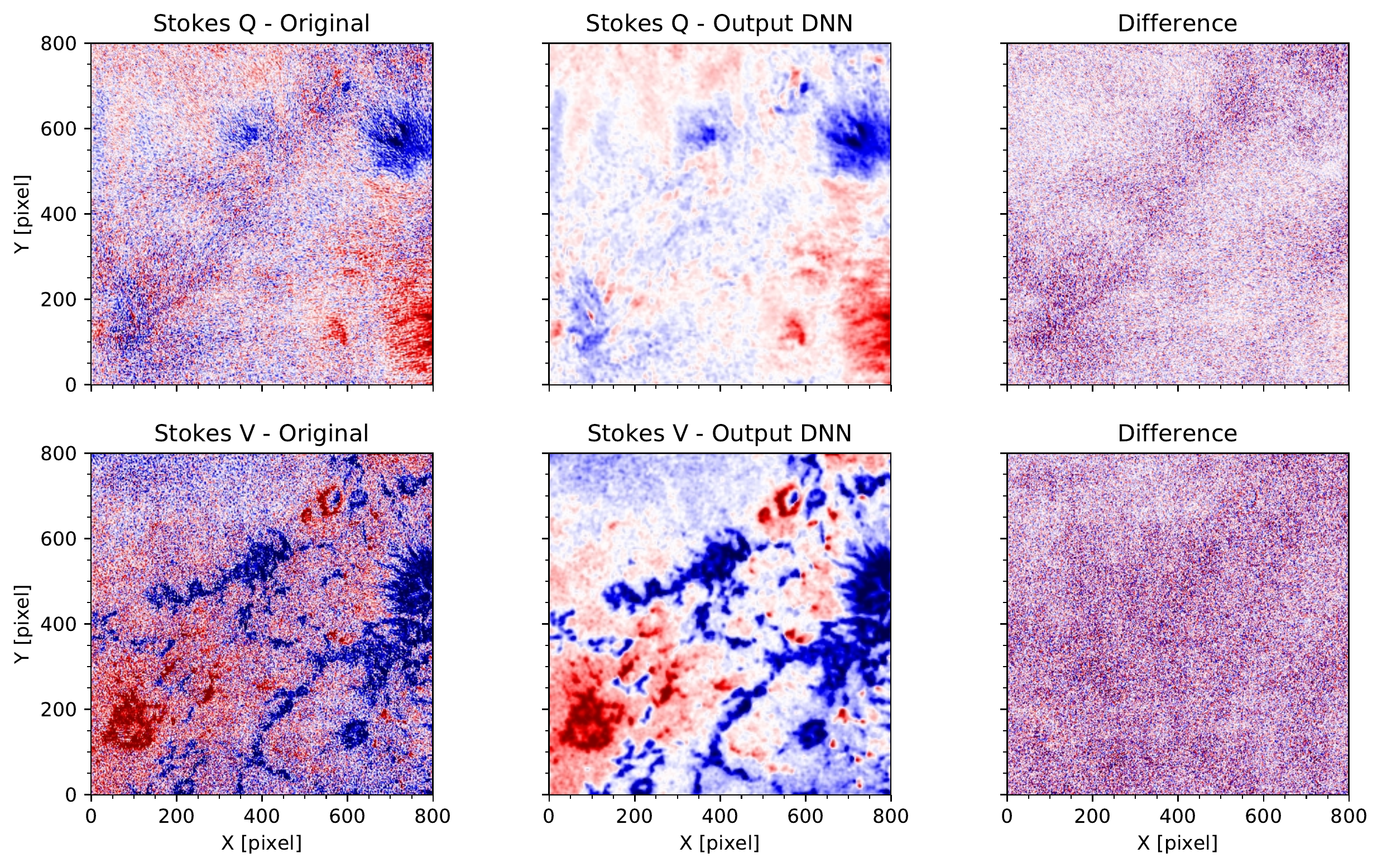}
\caption{Monochromatic maps of the result of the neural network applied to the same observations used for the training process.}
\label{fig:validation}
\end{figure*}

\end{appendix}

\end{document}